\documentclass[useAMS,usenatbib]{mn2e}

\usepackage[pdftex]{color}
\usepackage[pdftex]{graphicx}
\usepackage[draft]{hyperref}
\usepackage[fleqn]{amsmath}     

\newcommand{\kms}{km\,s$^{-1}$}

\newcommand{\eps}{\ensuremath{\epsilon}}
\newcommand{\qbar}{\ensuremath{\overline{q}}}

\title[A discrete chemo-dynamical model of Sculptor]{A discrete chemo-dynamical model of the dwarf spheroidal galaxy Sculptor: mass profile, velocity anisotropy and internal rotation}

\author[Zhu, van~de~Ven \& Watkins]{%
	Ling Zhu$^1$\thanks{E-mail: lzhu@mpia.de},
    	Glenn~van~de~Ven$^1$, Laura~L.~Watkins$^2$, Lorenzo Posti$^3$  \\
    	$^1$ Max Planck Institute for Astronomy, K\"onigstuhl 17, 69117 Heidelberg, Germany \\
    	$^2$ Space Telescope Science Institute, 3700 San Martin Drive, Baltimore, MD 21218, USA\\
    	$^3$ Kapteyn Astronomical Institute, University of Groningen, P.O. Box 800, 9700 AV Groningen, The Netherlands
	}

\date{Accepted 0000 Month 00. Received 0000 Month 00; in original 0000 Month 00}

\pagerange{\pageref{firstpage}--\pageref{lastpage}} \pubyear{2014}

\begin{document}

\label{firstpage}

\maketitle

\begin{abstract}
We present a new discrete chemo-dynamical axisymmetric modeling technique, which we apply to the dwarf spheroidal galaxy Sculptor. The major improvement over previous Jeans models is that realistic chemical distributions are included directly in the dynamical modelling of the discrete data. This avoids loss of information due to spatial binning and eliminates the need for hard cuts to remove contaminants and to separate stars based on their chemical properties. 
Using a combined likelihood in position, metallicity and kinematics, we find that our models naturally separate Sculptor stars into a metal-rich and a metal-poor population. 
Allowing for non-spherical symmetry, our approach provides a central slope of the dark matter density of $\gamma = 0.5 \pm 0.3$.
The metal-rich population is nearly isotropic (with $\beta_r^{red} = 0.0\pm0.1$) while the metal-poor population is tangentially anisotropic (with $\beta_r^{blue} = -0.2\pm0.1$) around the half light radius of $0.26$ kpc. A weak internal rotation of the metal-rich population is revealed with $v_{max}/\sigma_0 = 0.15 \pm 0.15$.
We run tests using mock data to show that a discrete dataset with $\sim 6000$ stars is required to distinguish between a core ($\gamma = 0$) and cusp ($\gamma = 1$), and to constrain the possible internal rotation to better than $1\,\sigma$ confidence with our model.
We conclude that our discrete chemo-dynamical modelling technique provides a flexible and powerful tool to robustly constrain the internal dynamics of multiple populations, and the total mass distribution in a stellar system.
\end{abstract}


\section{Introduction}
\label{S:intro}
Recent advances in both the quantity and quality of data available for the local population of dwarf spheroidal galaxies (dSphs) have revealed many complexities. The dSphs Carina, Fornax, Sculptor and Sextans all display evidence for the co-existence of at least two stellar populations: a spatially-concentrated metal-rich population and a spatially-extended metal-poor population (e.g. \citealt{Battaglia2008a}; \citealt{Kleyna2004}; \citealt{Tolstoy2004};  \citealt{Koch2008}). Moreover, the velocity dispersion profiles of the metal-poor stars are usually quite flat, while the metal-rich stars tend to have smaller velocity dispersion with profiles that decline sharply with radius (\citealt{Battaglia2008a};  \citealt{Battaglia2011}; \citealt{Amorisco2012Fornax}). This supports the idea that the dSphs have undergone at least two different star-formation episodes.

Study of the internal dynamics of these systems is crucial for understanding the mechanisms that drive their second epoch of star formation. The metal-rich secondary population may have formed from (re)accreted centrally-concentrated gaseous material. N-body simulations show that such a disk-like origin for a second-generation stellar population can leave behind significant kinematic signatures, including internal rotation, lower velocity dispersion, and velocity anisotropy (e.g., \citealt{AMB2013}). Violent processes, such as tidal stirring, collisions and mergers, can also compress the gas and trigger a second period of star formation, possibly resulting in counter-rotation or misalignment of the rotation axis (e.g., \citealt{Amorisco2012Fornax}).

Another major motivation for separating stars into multiple populations is to provide stronger constraints on the underlying gravitational potential of the system. There is still considerable debate as to whether the central regions of such halos are cored or cusped (e.g. \citealt{Kleyna2002}; \citealt{Koch2007}; \citealt{Walker2009jeans}; \citealt{Walker2011}; \citealt{Agnello2012}; \citealt{Amorisco2012Sculptor}; \citealt{Breddels2013}). dSphs provide an unprecedented opportunity to test the inner structure of dark matter (DM) halos as they are DM-dominated at all radii and they are dust free (e.g. \citealt{Mateo1998}; \citealt{Walker2013} and reference therein). The co-existence of multiple populations also enhances our ability to distinguish between a cored and cusped halos. 

In previous studies that have used a hard cut in metallicity to separate these different stellar populations (e.g. \citealt{Battaglia2008a}), the background contamination is hard to remove and the stars in the overlap region are excluded from either of the populations . Even so, some cross contamination will likely remain and significantly affect the dynamics of a population. Following a hard separation of the two populations, a few two-component dynamical studies were carried out to constrain the underlying DM halo by fitting the binned velocity dispersion profiles of the two populations simultaneously (e.g. \citealt{Amorisco2012Sculptor}; \citealt{Battaglia2008a}). 

There are a few studies that have separated multiple stellar populations using a combined likelihood for the spatial, chemical and velocity distributions to alleviate the cross contamination (e.g., \citealt{Amorisco2012Fornax}; \citealt{Walker2011}), and constrain the mass slope simultaneously with the virial mass estimates (e.g.,\citealt{Walker2011}). 
For example, \citet{Amorisco2012Fornax} showed that a division into three populations is preferred for Fornax. They found internal rotation in the intermediate and metal-rich populations and possible counter-rotation in the metal-poor population. As these rotation signals are very weak, they could never be found by separating multiple populations using a hard cut in metallicity. 
The virial masses for the two components obtained simultaneously gives a strong constraint on the DM slope (\citealt{Walker2011}; \citealt{Agnello2012}), they exclude a cusped DM halo with high significance. However, these studies assume that Sculptor is spherical, whereas it has been shown to have a flattening of $\sim 0.72$. When this ellipticity is accounted for, \citet{Walker2011} find that Sculptor prefers a density slope between a `core' and `cusp'. Another limit of this method is that virial estimates also assume that both populations are dynamically isotropic and do not allow for possible anisotropy.

In this paper, we construct chemo-dynamical models to separate multiple populations and fit a dynamical model simultaneously with different assumptions.
The discrete Jeans Anisotropic MGE (JAM) model \citep{Watkins2013} uses the velocity information on the observational plane, thus has the ability to recover the velocity anisotropy of the discrete system similar to that of the system with Integral Field Unit data (\citealt{Cappellari2008}; \citealt{Li2016}).
We extend the single-component discrete JAM models described in \citet{Watkins2013} to include multiple populations. We consider different stellar populations, tracing the same gravitational potential but each with its own spatial, chemical and dynamical distributions. The models are axisymmetric, the morphology of each population is free to be flattened, and each population has velocity dispersion and velocity anisotropy that follow the solution of the JAM model on the observational plane.  

We demonstrate the power of our modelling technique by applying it to several sets of mock data and the dSph galaxy Sculptor. Section~\ref{S:model} describes the model implementation and Section~\ref{S:mock} describes the application to the mock data.
In Section~\ref{S:Sculptor}, we apply it to the real data of Sculptor. We discuss our result in Section~\ref{S:discussion} and conclude in Section~\ref{S:conclusions}. In the appendix, we clarify the calculation of the first velocity moment in the Jeans models. 

\section{Discrete Chemo-dynamical models}
\label{S:model}

Consider a dataset of $N$ stars such that the $i$th star has sky coordinates $(x'_i, y'_i)$ and line-of-sight velocities $v_{z',i} \pm \delta{v_{z',i}}$. Here, $x'$ and $y'$ are along the projected major axis and minor axis, and $z'$ is along the line of sight. 
We follow the discrete dynamical modeling approach introduced by \citet{Watkins2013}, in this case, without proper motions (PMs) but with the addition of metallicities $Z_i\pm\delta Z_i$ in order to construct discrete chemo-dynamical models. 

We consider different stellar populations tracing the same gravitational potential, but each with its own chemical, spatial and dynamical properties. Even though the models can be generalised to have $k$ chemically-different stellar populations, in what follows, we assume $k=2$ populations, consisting of a metal-rich and a metal-poor stellar population, which we call red and blue populations.
Finally, contaminating or background stars are included as a third population with a uniform spatial density, and simple metallicity and dynamical distributions consistent with the Milky Way stellar halo. 

\subsection{Gravitational potential}
\label{SS:potential}

DSph galaxies like Sculptor typically have high mass-to-light ratios, up to $\sim$100 (e.g \citealt{Walker2013} and references therein), indicating that DM dominates at all scales and that the contributions from luminous matter can be neglected.
We adopt a generalised NFW (gNFW) density distribution
\begin{equation}
	\rho(r) = \frac{\rho_{s} }{ (r/r_s)^{\gamma}(1 + r/r_s)^{3-\gamma} }.
	\label{eq:densgNFW}
\end{equation}
In the axisymmetric case, $r^2 = x^2 + y^2 + z^2/q_h^2$, but since the flattening $q_h$ of the DM halo is, to a large degree, degenerate with its radial profile, line-of-sight data alone is expected to provide weak constraints if both are left free. Hence, for this first application of our new discrete chemo-dynamical model we will consider a spherical DM halo with $q_h = 1$. 

There are three free halo parameters: the scale radius $r_s$, the scale density $\rho_s$, and the inner density slope $\gamma$ in the potential.
When $\gamma =1$, this leads to a cusped profile, while, for $\gamma =0$, there is a core in the center.
In what follows, we will first leave $\gamma$ free,
we will consider later the two cases of a cusped $\gamma =1$ and cored $\gamma =0$ DM halos.

We use a Multi-Gaussian Expansion (MGE) of the density $\rho$ to simplify various calculations such as the computation of the gravitational potential \citep{Emsellem1994} and the solution of the axisymmetric Jeans equation \citep{Cappellari2008}. 

\subsection{Chemical probability}
\label{SS:probchm}

For each population $k$, we adopt a Gaussian distribution in metallicity with the mean metallicity $Z_0^k$ and metallicity dispersion $\sigma_Z^{k} $ as two free parameters. Given a star $i$ with measured metallicity  $Z_i\pm\delta{Z_i}$, the chemical probability for population $k$ is then
\begin{equation}
	\label{eq:pchm}
	P_{\mathrm{chm},i}^k = \frac{1}{\sqrt{2\pi[(\sigma_{Z}^k)^2+\delta{Z_i}^2]}}
		\exp\left[ -\frac12\frac{(Z_i-Z_0^k)^2}{[(\sigma_Z^k)^2 + \delta{Z_i}^2]}\right].
\end{equation}
%

\subsection{Spatial probability}
\label{SS:probspa}

Each population has its own spatial distribution through the observed surface number density $\Sigma^k(x',y')$.  Given a star $i$ at position $(x'_i,y'_i)$, the spatial probability for population $k$ is then
\begin{equation}
	\label{eq:pspa}
	P_{\mathrm{spa},i}^k  = \frac{\Sigma^k(x'_i,y'_i)}{\Sigma_\mathrm{obj}(x'_i,y'_i)+\Sigma_\mathrm{bg}},
\end{equation}
where $\Sigma_{\mathrm{obj}}$ is the combined surface number density of all populations that belong to the object under consideration, excluding the background surface number density $\Sigma_{\mathrm{bg}}$, which we assume to be uniform over the extent of the object and parameterize as a fraction $\eps$ of the central object surface number density, so that $\Sigma_{\mathrm{bg}} = \eps \, \Sigma_{\mathrm{obj}}(0,0)$.

Obtaining $\Sigma^k$ for each population can be challenging. In an optimal case, we construct the surface number density of all stars from a complete photometric catalogue, which we then expand into $M$ Gaussians. We then consider that each Gaussian $j$ contributes a fraction $h^k_j$ to the surface number density of each population so that
\begin{equation}
	\label{eq:sbcompmge}
	\Sigma^{k}(x', y') = \sum_{j=1}^M \, h^k_j \, \frac{L_j}{2\pi\sigma_j^2} 
	\exp\left[ -\frac{x'^2 + y'^2/{q'}_j^2}{2\sigma_j^2}\right],
\end{equation}
where $L_j$, $\sigma_j$, ${q'}_j$ are the total luminosity, dispersion and projected flattening of each Gaussian $j$. 

If we impose the constraint $\sum_k h^k_j = 1$, then this implies, in the case of the two stellar populations considered here, that if $h^\mathrm{red} = h_j$ then $h^\mathrm{blue} = 1 - h_j$ for every Gaussian.
The fractions $h_j$ will be constrained through both the spatial and dynamical probability, because the velocity distribution for a given tracer population predicted by a dynamical model depends on both the gravitational potential and the surface number density of the tracer population.

\subsection{Dynamical probability}
\label{SS:probdyn}

Given a star $i$ with measured line-of-sight velocity $v_{z',i} \pm \delta{v_{z',i}}$, the dynamical probability for population $k$ for an assumed Gaussian velocity distribution is then
\begin{equation}
	\label{eq:pdyn}
	P_{\mathrm{dyn},i}^k  = \frac{1}{\sqrt{(\sigma^k_i)^2 + (\delta v_{z',i})^2}}
	\exp\left[ -\frac12\frac{(v_{z', i} - \mu^k_{i})^2 }{(\sigma^k_i)^2 + (\delta v_{z',i})^2}\right],
\end{equation}
where $\mu_i^k$ and $\sigma_i^k$ are the line-of-sight mean velocity and velocity dispersion as predicted by a dynamical model at the sky position $(x'_i,y'_i)$.

Following Section~4 of \citet{Watkins2013}  \citep[see also][]{Cappellari2008}, we adopt here as a dynamical model the solution of the axisymmetric Jeans equations under the two assumptions that: (i) the velocity ellipsoid is aligned with the cylindrical coordinate system so that $\overline{v_R v_z}=0$; and (ii) the velocity anisotropy in the meridional plane $\beta^k_z = 1 - \overline{v_z^2} / \overline{v_R^2}$ is constant. When the gravitational potential and tracer density are expressed in terms of an MGE, as in our case, the solution for the second-order velocity moments reduces to a single numerical integral; this includes the integration along the line-of-sight for a given inclination $\vartheta$ at which the object is observed.
The first-order velocity moments follow after setting the relative contribution of ordered and random motions via a rotation parameter $\kappa^k$ (\citealt{Cappellari2008}, but see Appendix \ref{S:Arotation} for our clarified definition of rotation).  
The combination of predicted first-order and second-order velocity moments yields $\mu_i^k$ and $\sigma_i^k$ for each population $k$ as position $(x', y')$.

Since the anisotropy parameter $\beta^k_{z,j}$ and rotation parameter $\kappa^k_j$ for each Gaussian $j$ in the MGE of the tracer density can, in principle, take on a different constant value, it is possible to model velocity anisotropy in the meridional plane and intrinsic rotation that vary with radius. However, in the current analysis, to restrict the number of free parameters, we adopt a radially constant anisotropy and rotation, i.e., $\beta^k_{z,j} = \beta^k_z $ and  $\kappa^k_j = \kappa^k$ for all Gaussians $j$. However $\beta^k_z$ and $\kappa^k$ are allowed to vary between different populations $k$.

\subsection{Background}
\label{SS:background}

The main contamination comes from Milky Way halo stars. 
We adopt a Gaussian metallicity distribution with fixed mean $Z_{0}^\mathrm{bg}$ and $\sigma_Z^\mathrm{bg}$.   
As mentioned in Section~\ref{SS:probspa}, the background surface number density is assumed to be uniform across the extent of an extragalactic stellar object, with free parameter $\eps$ accounting for the level relative to the central surface number density of the object.
Finally, the velocity distribution is assumed to be Gaussian with mean velocity $\mu_0^\mathrm{bg} = -V_\mathrm{sys}$, the systematic velocity of the object compared to the Milky Way stellar halo, and dispersion $\sigma_0^\mathrm{bg}$ also fixed.

\subsection{Total probability}
\label{SS:probtotal}

Combining the above chemical, spatial and dynamical probabilities, the likelihood for star $i$ is
\begin{multline}
	\label{eq:likelihood_i}
	L_i =  \sum_{k\ne\mathrm{bg}} P_{\mathrm{spa},i}^k \,P_{\mathrm{chm},i}^k \, P_{\mathrm{dyn},i}^k 
	\\
	+ \left( 1 - \sum_{k\ne\mathrm{bg}} P_{\mathrm{spa},i}^k \right) \, P_{\mathrm{chm},i}^\mathrm{bg} \, P_{\mathrm{dyn},i}^\mathrm{bg}.
\end{multline}
The summation is over all populations that belong to the object under consideration; in the current study, this is a red and blue population, in addition to the background. 
The total likelihood $L = \prod_{i=1}^{N} L_i$ of all $N$ stars is the quantity we wish to maximise.

For a model with all parameters known, the likelihood of each star $i$ to be within each population $k$ is
\begin{equation}
P_i^k = P_{spa,i}^k P_{chm,i}^k P_{dyn,i}^k,
\end{equation}
where $k$ can be red, blue or the background.
Then, the relative value
\begin{equation}
\label{eq:Pik}
P_i^{'k} = P_i^k / \sum^k P_i^k
\end{equation}
can be used to identify the stars to be red, blue or background stars separated by this model.

\subsection{Model parameters}
\label{SS:parameters}

Here we summarise the free parameters in our discrete chemo-dynamical model of a stellar system with two chemically distinct stellar populations.
 

Under the assumption that the gravitational potential is dominated by a spherical DM halo with generalised NFW radial mass density profile, there are 3 free \emph{potential} parameters:
\begin{itemize}
\item[(1)] $\rho_s$, the scale density;
\item[(2)] $r_s$, the scale density; 
\item[(3)] $\gamma$, central density slope: cusped $\gamma =1$ vs.\ cored $\gamma =0$;
\end{itemize}
Under the assumption that the distribution of stars in the object under study is oblate axisymmetric and that, over its extent, the distribution of contaminating stars is uniform, the viewing orientation and background influence are described with 2 free \emph{global} parameters:
\begin{itemize}
\item[(4)] $\qbar$, average intrinsic flattening, directly related inclination angle $\vartheta$ via the relation $\overline{q'}^2 = \cos\vartheta^2 + \qbar^2 \sin\vartheta^2$, given the observed average flattening $\overline{q'}$;
\item[(5)] $\eps$, fraction background surface number density level relative to center of object, so that $\Sigma_\mathrm{bg} = \eps \, \Sigma_\mathrm{obj}(0,0)$;
\end{itemize}
We further assume that the stellar system consists of a red (metal-rich) and blue (metal-poor) stellar populations, both with a Gaussian metallicity distribution and a Gaussian line-of-sight velocity distribution as predicted by an axisymmetric Jeans model, this adds 4 free \emph{population} parameters per population. For the red population, they are:
\begin{itemize}
\item[(6)]  $Z_0^\mathrm{red}$, mean of the Gaussian metallicity distribution for the red population;
\item[(7)] $\sigma_Z^\mathrm{red}$, dispersion of the Gaussian metallicity distribution;
\item[(8)] $\lambda^\mathrm{red} \equiv - \ln \left( 1 - \beta^\mathrm{red}_z \right)$ , symmetric re-casting the constant velocity anisotropy in the meridional plane $\beta^\mathrm{red}_z$;
\item[(9)] $\kappa^\mathrm{red}$ , rotation parameter for the red population;
\end{itemize}
Correspondingly, for the blue population, the free parameters are:
\begin{itemize}
\item[(10)]  $Z_0^\mathrm{blue}$;
\item[(11)] $\sigma_Z^\mathrm{blue}$;
\item[(12)] $\lambda^\mathrm{blue}$; 
\item[(13)] $\kappa^\mathrm{blue} $; 
\end{itemize}
Finally, depending on the number of Gaussian components into which the observed total surface number density distribution is being decomposed, there will be additional parameters $h_j$, describing the fractional contribution of the red stellar population to each of the Gaussian components, that are left free.

\section{Application to mock data}
\label{S:mock}
The ability of axisymmetric Jeans models to recover the mass profile and velocity anisotropy of different types of galaxies has already been established statistically with thousands of simulated galaxies in \citet{Li2016}. We focus on testing how well we are able to distinguish two chemically and kinematically distinct populations from discrete data, while simultaneously recovering the underlying gravitational potential.  

Two-component mock data sets are available at the Gaia Challenge wiki \footnote{ astrowiki.ph.surrey.ac.uk/dokuwiki/doku.php?id=tests:sphtri}, but they are all spherically symmetric. 
Therefore, we create our own axisymmetric two-component mock data sets using analytic distribution functions, and make them public as online material with this paper. 
The mock data sets have been generated using a modified version of the 
\textsc{Agama}\footnote{\url{https://github.com/GalacticDynamics-Oxford/Agama}}
code, whose functionalities will be described in an upcoming paper (Vasiliev et al. in preparation).

\subsection{Mock data}
\label{SS:mockdata}
\begin{figure}
\centering\includegraphics[width=4.5cm]{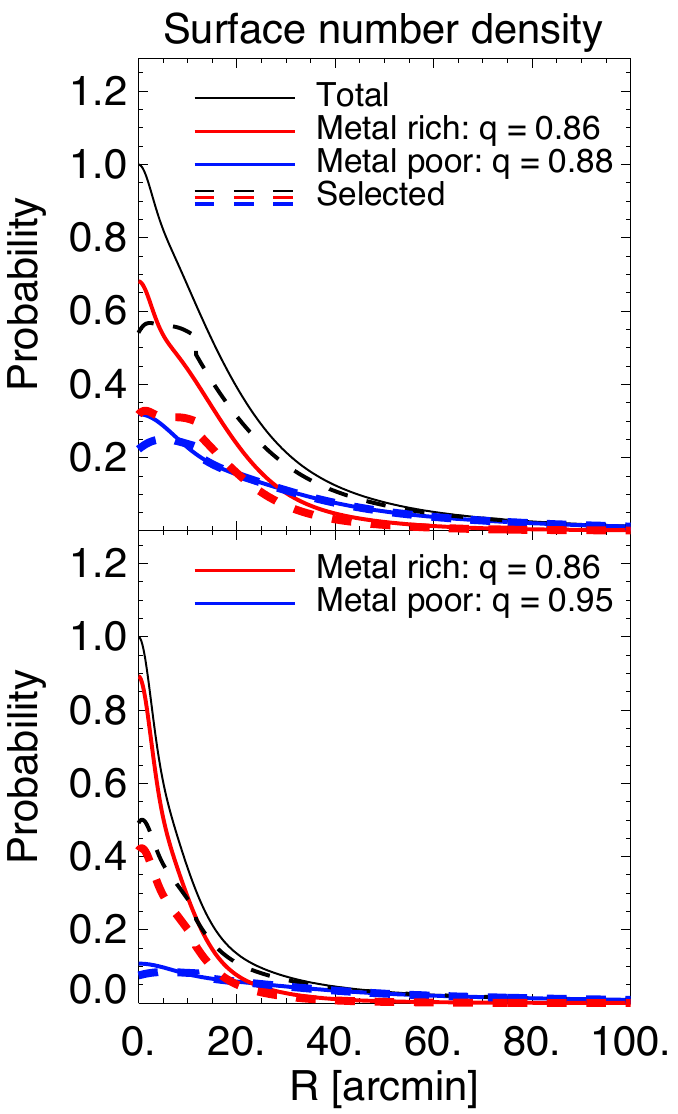}\centering\includegraphics[width=3.96cm]{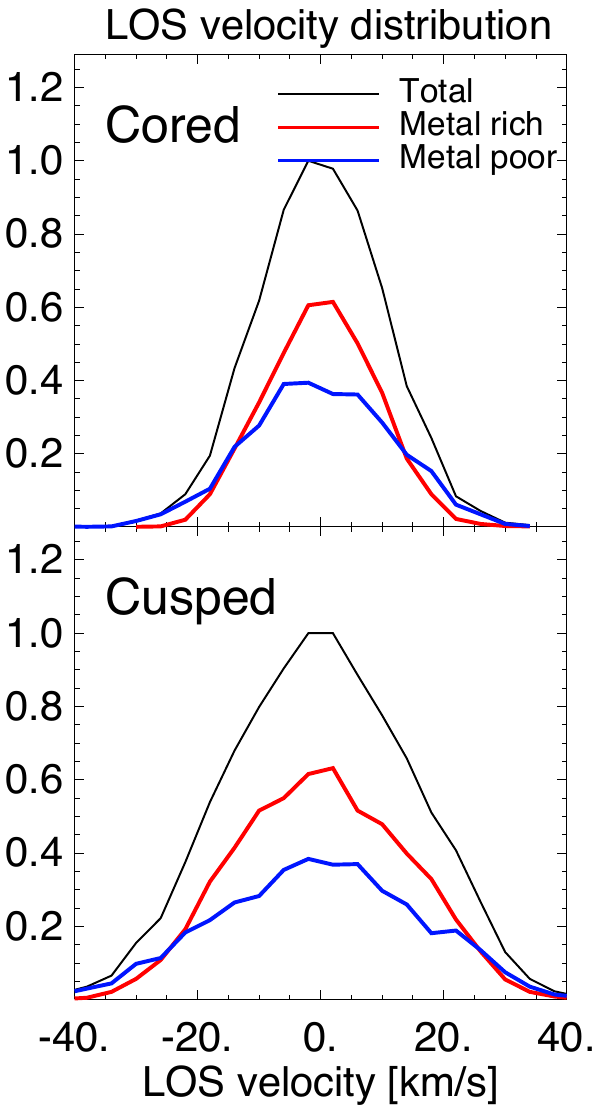}
\caption{The surface number density profiles and the velocity distributions of the mock data. The upper panels are for the cored mock data and the bottom panels are for the cusped mock data. Red lines represent metal-rich populations, blue lines represent metal-poor populations, and the black lines are the total. In the left panels, the solid lines represent the original surface number density profiles and the dashed lines are those for the selected discrete data points.}
\label{fig:vsb_dis}
\end{figure}

\begin{figure}
\centering\includegraphics[width=\hsize]{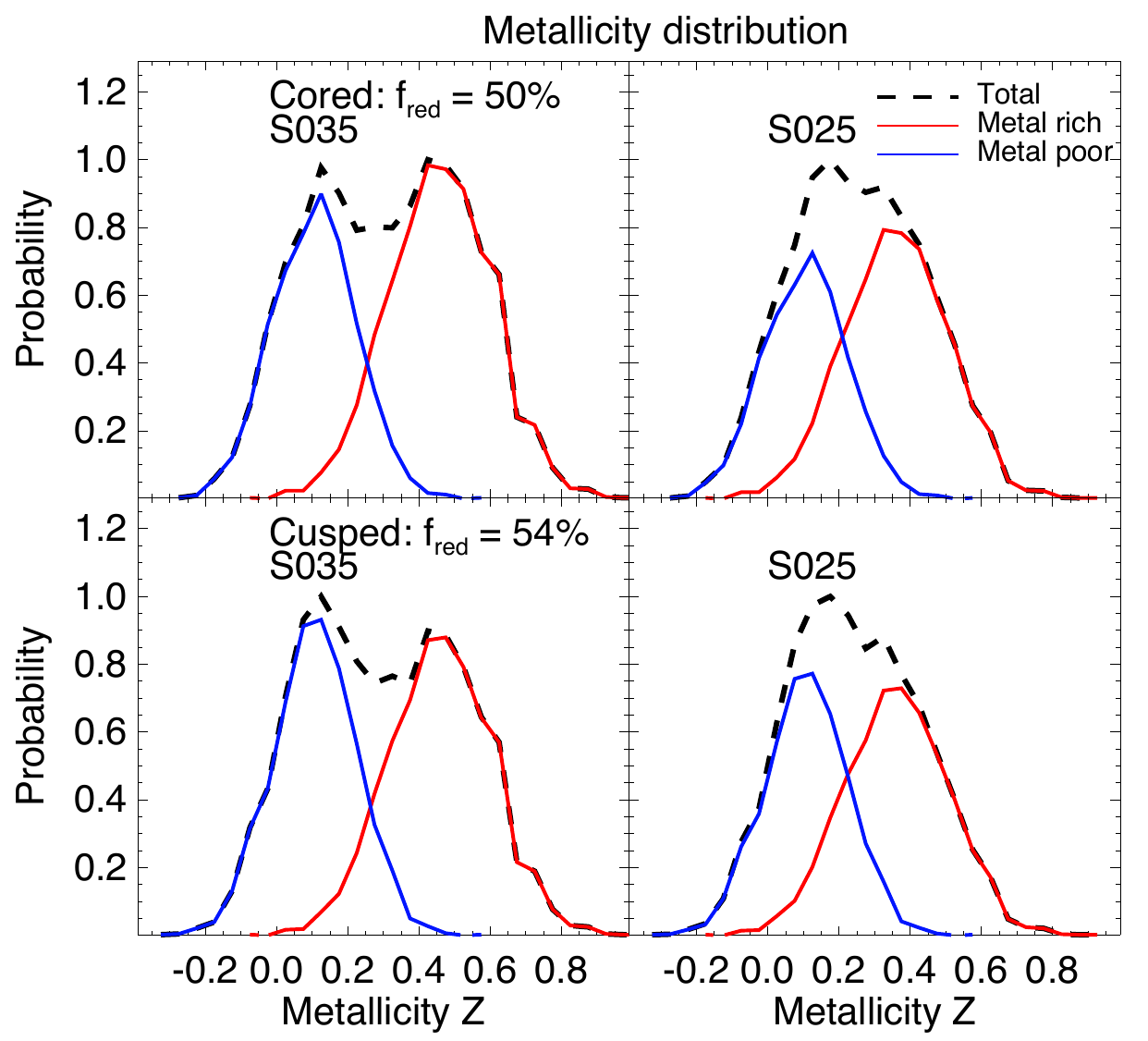}
\caption{The metallicity distributions. $f_{\mathrm{red}}$ indicates the fraction of red stars in the sample. We create two metallicity distributions: \emph{S035} with 0.35 dex separation and \emph{S025} with 0.25 dex separation of the two populations. The metallicity separation of the two populations in Sculptor is similar to \emph{S025}. }
\label{fig:cr_dis}
\end{figure}

We create mock data as follows:

\begin{itemize}
\item[-] Gravitational potential: We adopt a generalized NFW profile (see equation~\ref{eq:densgNFW}) for the DM halo and generate two sets of mock data: one with a cusped halo ($\gamma = 1$) and one with a cored halo ($\gamma = 0$). The input gravitational potential parameters are listed in Table~\ref{tab:mock}. 

\item[-] Tracer densities and kinematics: In general, we generate two stellar components
as discrete realizations of continuous models whose distribution functions
(DFs) are analytic functions of three action integrals $\bf J$. By sampling
the DFs we self-consistently generate both the positions and velocities
of a sample of 5000 stars for each population under the influence of the DM halo potential.
We employ $f=f({\bf J})$ DFs that are double power-laws in the action
integrals, and so generate stellar density distributions that are double
power-laws of radius. In particular, we fix the two power-law slopes so
to have stellar density distributions which closely follow Isochrone
\citep{Henon1959} distributions (see \citealt{Posti2015}, Section 4.1). The
models are axisymmetric, anisotropic, rotating and flattened (either
by anisotropy or rotation) and are described in Posti et al. (in prep)
who introduced them to model elliptical and lenticular galaxies. 

The spatial distributions and kinematics of the two stellar populations are similar but not identical in the two mock data sets with different DM halos.

In both mock data sets, the metal-rich population
has an Isochrone scale radius that is approximately twice as small as that for the metal-poor population.

The metal-rich population has a non-negligible rotation velocity peaking at about $\sim 9$km/s. It is almost
isotropic within the constant-density core, and becomes tangentially-biased further out.
The metal-poor population is effectively
non-rotating. The velocity anisotropy is different in the two mock data sets: in the cusped mock data, the metal-poor population has similar velocity anisotropy to the metal-rich population, while in the cored mock data it becomes radially-biased further out. The overall velocity distributions are shown in the right panels of Figure~\ref{fig:vsb_dis}.

\item[-] Metallicity: We assume that the metallicity distribution of each population follows a Gaussian profile, with different separations of the two populations. We create two metallicity distributions: \emph{S035} with 0.35 dex separation and \emph{S025} with 0.25 dex separation of the two populations, for each set of the data. 
The combined metallicity distributions of our samples are shown in Figure~\ref{fig:cr_dis}, with the corresponding Gaussian parameters in Table~\ref{tab:mock}. 

\item[-] Simulate real data: We project the system $20^o$ away from edge-on with inclination angle $\vartheta = 70^{\circ}$, and place it at a distance of 79 kpc. Then we extract the position, line-of-sight velocity, and metallicity of each star. We randomly draw points with ($x_i$, $y_i$, $v_{z,i} + \delta v_{z,i}$, $Z_{i} + \delta Z_i$) from the blue population and red population. The spatial sampling is biased by the selection function that stars in the inner regions are less likely selected \citep{Walker2011}, and we truncate at the projected radius of 2 kpc. The surface number density profile of the selected discrete data points (dashed lines) are shown in Fig~\ref{fig:vsb_dis}. The velocity and metallicity have been perturbed with typical velocity error of 3 \kms and typical metallicity error of $0.1$ dex. 

\item[-] Combine populations: we combine the data points from the two populations together, yielding a red population fraction of $f_{red} = 50\%$ in the cored samples and $f_{red} = 54\%$ in the cusped samples.  

\item[-] Sample selection: for the \emph{S025} samples, we draw $1/3$ of the stars at random to form a new sample, which we refer to as the \emph{S035 1/3} sample. With $\sim 2000$ stars, the so called \emph{S025 1/3} samples have similar size of the real data sample we have for Sculptor. The kinematic properties and metallicity distributions of this $1/3$ sample are kept the same as the corresponding full sample. We create six mock data sets in total. 
\end{itemize}

The six mock data sets are summarized in Table~\ref{tab:mock}. The velocity anisotropy parameters ($\lambda^\mathrm{blue}_z$, $\lambda^\mathrm{red}_z$) and rotation parameters ($\kappa^{blue}$, $\kappa^{red}$) are calculated from the mock data with full 6D ($x,y,z,v_x,v_y,v_z$) information. They actually vary with radius in the mock data, here we only show the average values.

\begin{table*}
\caption{The input parameters for the mock data, from left to right: name of the sample, sample size, fraction of the red population $f_{red}$, DM scale density $\rho_s\, [M_{\sun} /\mathrm{pc}^3]$ , the scale radius $r_s \, [\mathrm{pc}]$, central density slope $\gamma$, the intrinsic flattening for the metal-poor $\overline{q}^{\mathrm{blue}}$ and metal-rich $\overline{q}^{\mathrm{red}}$ populations, mean metallicity $Z_0^\mathrm{blue}$ and $Z_0^\mathrm{red}$,  metallicity spread $\sigma_Z^\mathrm{blue}$ and  $\sigma_Z^\mathrm{red}$, 
velocity anisotropy parameters $\lambda^\mathrm{blue}_z$ and $\lambda^\mathrm{red}_z$, the rotation parameters $\kappa^{blue}$ and $\kappa^{red}$. The velocity anisotropy parameters ($\lambda^\mathrm{blue}_z$, $\lambda^\mathrm{red}_z$) and rotation parameters ($\kappa^{blue}$, $\kappa^{red}$) vary with radius in the mock data, here we only show the average values.}
\label{tab:mock}
\small
\scriptsize
\begin{tabular}{llllllllllllllllll}
\hline
\hline
 Data & sample size & $f^{\mathrm{red}}$ $\%$ &$\log(\rho_s)$ & $\log(r_s)$ & $\gamma$ & $\overline{q}^{\mathrm{blue}}$ & $\overline{q}^{\mathrm{red}}$  &  $Z_0^\mathrm{blue}$ & $Z_0^\mathrm{red}$  & $\sigma_Z^\mathrm{blue}$  & $\sigma_Z^\mathrm{red}$ & $\lambda_z^\mathrm{blue}$ &  $\lambda_z^\mathrm{red}$ &  $\kappa ^{blue}$  & $\kappa^{red}$    \\
  \hline 
  \hline
  Cored & & & & & & & & & & & & & & & &\\
\emph{S035}   &  6417 & 50&  -0.189 & 3.0 &  0.0 & 0.88 & 0.86  & 0.10  & 0.45  &  0.12 & 0.15 & 0.4 & -0.1 & 0.00 & 0.35  \\
\emph{S025}   & -  &  - & - &  - &  -  & - & -& 0.10  & 0.35  &  - & -& - & - & - & -   \\
\emph{S025 1/3}  & 2250  &  - & - &  - &  -  & - & -& 0.10  & 0.35  &  - & -& - & - & - & -   \\
  Cusped & & & & & & & & & & & & & & &\\
\emph{S035}   &  5360 & 54 & -1.189  & 3.0 &  1.0 & 0.95 & 0.86   & 0.10  & 0.45  &  0.12 & 0.15 &  -0.1 & -0.1 & 0.00 & 0.30  \\
\emph{S025}   & -  &  - & - &  - &  -  & - & - & 0.10  & 0.35  &  - & -& - & - & - & -   \\
\emph{S025 1/3}  & 1610  &  - & - &  - &  -  & - & -& 0.10  & 0.35  &  - & -& - & - & - & -   \\
  \hline
  \hline
 \end{tabular}
\end{table*}

Our dynamical models require a surface number density profile for each tracer population in the form of an MGE. 
We know the surface number density profiles of the two populations in the mock data, however we do not use them in our chemo-dynamical model. With the real photometric data, the surface number densities of the true red and blue populations are usually unknown, so here we take a process similar to that used for real Sculptor data. 
The true surface number density profile of the red ($\Sigma^\mathrm{true\, red}(x',y')$) and blue population ($\Sigma^\mathrm{true\, blue}(x',y')$) will be taken as two backbone shapes; the surface number density profile for the red and blue population we put in the model will be a combination of these two shapes:
\begin{equation}
	\label{eq:numdenpop1}
	\Sigma^\mathrm{red}(x',y') = h_1 \, \Sigma^\mathrm{true\, red}(x',y') + h_2 \, \Sigma^\mathrm{true\, blue}(x',y'),
\end{equation}
and similarly for the blue population with fractions $1-h_1$ and $1-h_2$. As a result, the combined surface number density profile of the two populations is fixed, the fractions $h_1$ and $h_2$ are two free \emph{density} parameters in addition to the 13 free parameters described in Section~\ref{SS:parameters}.
We did not include background stars in our mock data, thus the background fraction $\eps$ is kept at zero. We have 14 free parameters in total when modelling the mock data. 

\subsection{Modelling}
\label{SS:modelling}

\subsubsection{The MCMC process}
\label{SS:mock_mcmc}

We apply our discrete chemo-dynamical model with a generalized NFW DM halo to the six mock data sets.  
We use the \textsc{emcee} package \citep{Foreman-Mackey2013} -- a pure Python implementation of the affine-invariant MCMC ensemble sampler -- to efficiently explore the parameter space of our models. For each set of models, 200 walkers with 600 steps are used. As members of the ensemble, the walkers are almost like separate Metropolis-Hasting chains except that the proposal distribution for a given walker depends on the positions of all the other walkers in the ensemble.

\begin{figure*}
\centering\includegraphics[width=\hsize]{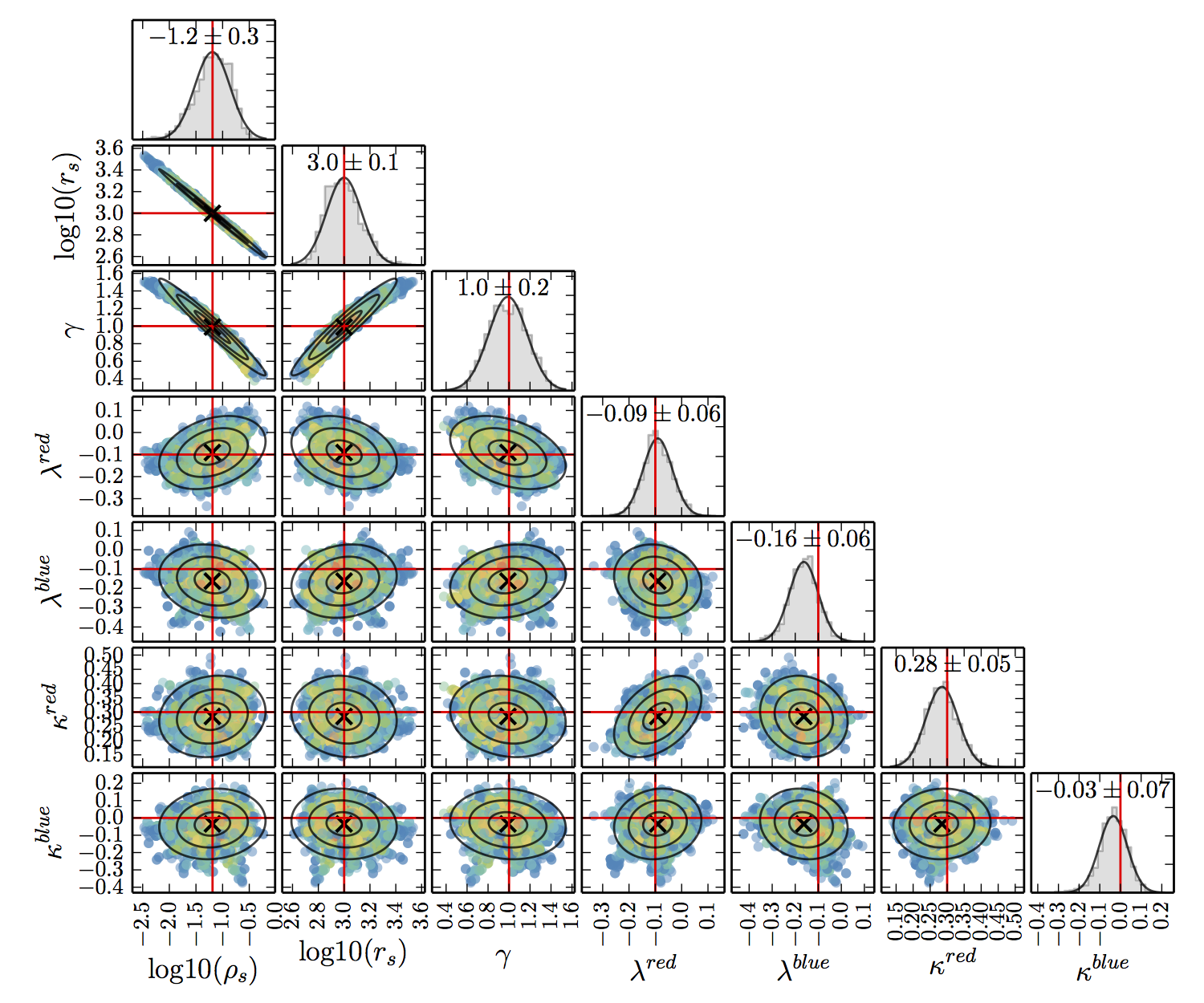}
\caption{MCMC post-burn distributions for a gNFW model of the cusped mock \emph{S035} dataset. The scatter plots show the projected two-dimensional distributions, with the points coloured by their likelihoods from blue (low) to red (high). The ellipses represent the $1\sigma$, $2\sigma$ and $3\sigma$ regions of the projected covariance matrix. The histograms show the projected one-dimensional distributions. The red lines represent the true values of the mock data. The parameters from left to right: DM scale density $\rho_s$, DM scale radius $r_s$, the DM inner density slope $\gamma$, velocity anisotropy parameter of the red $\lambda^\mathrm{red}$ and the blue population $\lambda^\mathrm{blue}$, rotation parameter of the red population $\kappa^\mathrm{red}$ and the blue population $\kappa^\mathrm{blue}$. }
\label{fig:mcmc_postburn}
\end{figure*}

The models converge well even with 14 free parameters. Figure~\ref{fig:mcmc_postburn} shows the projected two-dimensional distributions for 7 of the 14 parameters (omitting the parameters which are not directly related to the dynamics) for the gNFW model of the cusped mock \emph{S035} data set. 
 The red lines represent the true values of the mock data, and the black ellipses represent the $1\sigma$, $2\,\sigma$ and $3\,\sigma$ regions of the projected covariance matrix.
With the cusped \emph{S035} samples, our models recover the three potential parameters and the kinematic properties of the two populations perfectly well, although there is a strong degeneracy among the three potential parameters. 

The potential and the kinematic parameters are generally recovered by all the data sets, although as expected, the uncertainties increase with decreasing metallicity separations of the two populations (\emph{S025} samples) and with decreasing number of data points (\emph{S025 1/3} samples). The recovered model parameters for all the six mock data sets are summarized in Table ~\ref{tab:mock2}.

\begin{table*}
\caption{The parameters recovered by our chemo-dynamical model with different mock data sets. The general true values for the Cored and Cusped model are shown for comparison purpose. The parameters are presented in two rows for each model. The first column is the name of the sample, the second column marked the number of data points used, the first row following from left to right:  DM scale density $\rho_s$, DM scale radius $r_s$, central density slope $\gamma$, the inclination angle $\vartheta$, the fraction $h_1$ and $h_2$ of the true red and true blue surface number density profiles contributing to the red population.
Second row from left to right: mean metallicity $Z_0^\mathrm{blue}$ and $Z_0^\mathrm{red}$,  metallicity spread $\sigma_Z^\mathrm{blue}$ and  $\sigma_Z^\mathrm{red}$, 
velocity anisotropy parameter  $\lambda^\mathrm{blue}_z$ and $\lambda^\mathrm{red}_z$,  the rotation parameter $\kappa^{blue}$ and  $\kappa^{red}$. }
\label{tab:mock2}
\small
\scriptsize
\begin{tabular}{lllllllllllllllll}
\hline
\hline

 Data & points &$\log(\rho_s [M_{\sun} /\mathrm{pc}^3])$ & $\log(r_s  [\mathrm{pc}])$ & $\gamma$ & $\vartheta$ & $h_1$ & $h_2$  & \\
$Z_0^\mathrm{blue}$ & $Z_0^\mathrm{red}$ &   $\sigma_Z^\mathrm{blue}$  & $\sigma_Z^\mathrm{red}$   & $\lambda_z^\mathrm{blue}$ &  $\lambda_z^\mathrm{red}$ &  $\kappa ^{blue}$  & $\kappa^{red}$\\
\hline
\hline

Cored &$ - $&$   -1.2 $&$  3.0  $&$ 0 $&$ 70^{\circ}  $&$  1.0 $&$ 0.0  $ & \\
$ 0.1  $&$  0.45/0.35  $&$   0.12  $&$  0.15  $&$ 0.40 $&$ -0.1 $&$  0.0 $&$ 0.35 $  \\    
   
\hline
\emph{S035} &$ 6417 $&$   -1.1\pm 0.2 $&$  2.97 \pm0.09  $&$ 0.1\pm0.2 $&$ 78^{\circ} \pm9  $&$  0.99\pm0.01 $&$ 0.02\pm0.01  $ & \\
$ 0.102 \pm 0.003  $&$  0.452 \pm0.009  $&$   0.117\pm0.002  $&$  0.141\pm0.006  $&$ 0.20 \pm0.05 $&$ -0.26 \pm 0.08 $&$  0.1\pm0.1 $&$ 0.32 \pm0.05 $  \\    

\emph{S025} &$ 6417 $&$  -1.3\pm  0.3 $&$  3.1\pm0.1   $&$ 0.3\pm0.3 $&$ 78^{\circ} \pm 9   $&$  0.98\pm  0.02 $&$ 0.03\pm0.02  $ & \\
 $ 0.106 \pm 0.004  $&$  0.363  \pm0.006   $&$   0.117\pm0.004  $&$  0.141\pm0.005  $&$ 0.3\pm 0.1 $&$ -0.4  \pm  0.1 $&$  0.1 \pm0.1 $&$ 0.32 \pm0.06 $  \\    
\emph{S025 1/3} &$ 2250 $&$  -1.5\pm  0.4 $&$  3.2 \pm0.2    $&$ 0.3\pm0.4 $&$ 79^{\circ} \pm 9  $&$  0.96\pm  0.03 $&$ 0.09\pm0.05   $ & \\ 
$ 0.105 \pm 0.008  $&$  0.38  \pm0.01   $&$   0.119\pm0.008  $&$  0.135\pm 0.01   $&$ 0.2 \pm 0.1  $&$ -0.3  \pm  0.2 $&$  0.0 \pm 0.2 $&$ 0.3 \pm0.1  $  \\

\hline\hline
Cusped&$ - $&$   -1.2  $&$ 3.0    $&$ 1  $&$  70^{\circ} $&$ 1.0 $&$ 0.0  $& \\
$  0.1  $&$  0.45/0.35  $&$  0.12 $&$ 0.15 $&$ -0.1 $&$ -0.1  $&$   0.0  $&$ 0.3$ \\   

\hline 
\emph{S035} &$ 5360 $&$   -1.2 \pm 0.3 $&$ 3.0 \pm0.1   $&$ 1.0\pm 0.2  $&$  78^{\circ}\pm 9 $&$ 0.95 \pm 0.01 $&$ 0.01 \pm 0.01  $& \\
$  0.105 \pm 0.004  $&$  0.460 \pm 0.004  $&$  0.118 \pm 0.005 $&$ 0.144 \pm 0.004 $&$ -0.16\pm 0.06$&$ -0.09\pm0.06  $&$   -0.03 \pm 0.07  $&$ 0.28\pm 0.05$ \\ 

\emph{S025} &$ 5360 $&$   -0.8\pm  0.4 $&$ 2.8\pm 0.2   $&$ 0.7\pm 0.3  $&$  79^{\circ} \pm 9 $&$ 0.95 \pm 0.02 $&$ 0.02 \pm 0.01  $& \\
$  0.104 \pm 0.005  $&$  0.362 \pm 0.004 $&$   0.118 \pm 0.005 $&$ 0.142 \pm 0.004  $&$ -0.16 \pm0.07 $&$  -0.07\pm0.05 $&$  -0.00\pm 0.07  $&$ 0.29\pm0.05$  \\ 
\emph{S025} 1/3 &$ 1610 $&$   -0.8\pm  0.6 $&$ 2.8\pm 0.2  $&$ 0.8\pm 0.4  $&$  79^{\circ}\pm  10 $&$  0.95\pm  0.03$&$ 0.05 \pm 0.03 $& \\
$  0.104 \pm 0.008    $&$ 0.371  \pm 0.008   $&$  0.119 \pm  0.009 $&$  0.138\pm  0.009 $&$  0.0 \pm 0.1 $&$  -0.2\pm  0.2 $&$   0.1\pm  0.3   $&$ 0.19\pm  0.08  $\\

  \hline
  \hline
 \end{tabular}
\end{table*}

\subsubsection{Probability distribution}
\label{SS:prob}
\begin{figure}
\centering\includegraphics[width=\hsize]{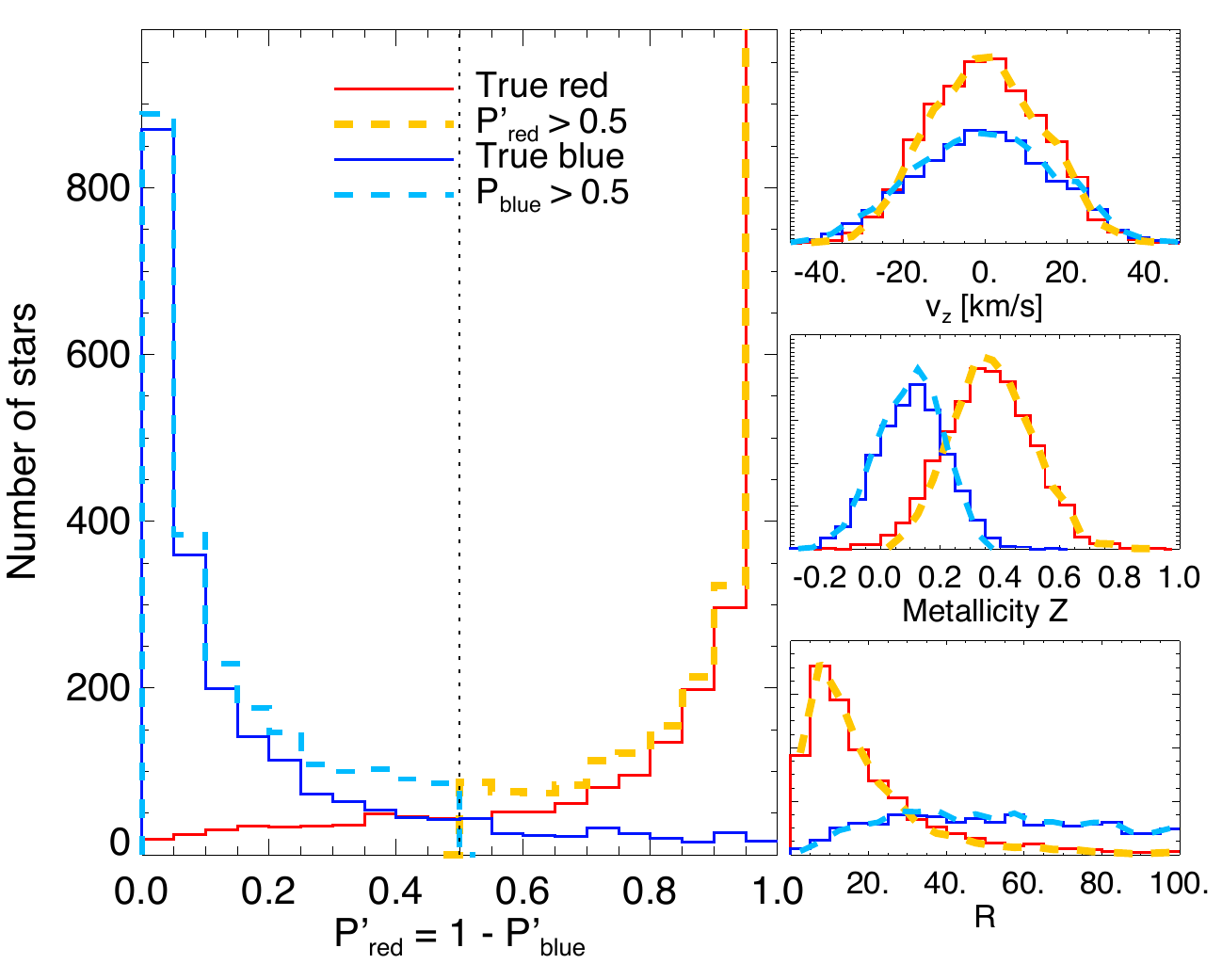}
\caption{The left panel shows the probability distribution of the stars in the best-fitting model of cusped \emph{S035} data. The red and blue solid lines represent the true red and true blue stars, the yellow dashed is for the model-identified red stars ($P_i^{'red} > 0.5$) and light blue dashed line is for the model-identified blue stars ($P_i^{'blue}>0.5$). The right panels show the corresponding spatial, metallicity and global velocity distribution of these four groups. The true red (blue) stars have similar distributions to the red (blue) stars identified by a probability cut.}
\label{fig:prob}
\end{figure}

Each star has its non-zero probability of belonging to the red or blue population in the model as we described in Section~\ref{SS:probtotal}. The probability distribution of stars in the best-fitting model of the cusped \emph{S035} data is shown in the left panel of Figure~\ref{fig:prob}.
The probability distribution of the true red stars peaked at $P'_{red} \sim 1$ as expected, then gradually decreases and has a long tail to $P'_{red} \sim 0$; the blue stars show a similar trend. 

With mock data, we know the true red and blue stars, so we can calculate the true kinematics of the two populations using their member stars and compare with our model prediction. However, in the real case, we do not know the true members of each population. Thus we want to find a way to extract the true kinematics of the two populations from the stars based on their probability distributions. 

In principle, we can extract properties of the red (blue) populations from all the stars weighted with $P'_{red}$ ($P'_{blue}$). Practically, we find this approach works well to get the kinematics of each population, but it tends to smooth the kinematical profiles, thus does not represent the fluctuations of the data well. 

As illustrated in Figure~\ref{fig:prob}, we find that if we take a probability cut at 0.5 ($P_i^{'red}>0.5$ for red; $P_i^{'blue}>0.5$ for blue) to separate the two populations, the majority of the red and blue stars will be identified correctly. Only a small fraction of stars (the long tails of the red and blue solid histograms) will be misidentified. The right panels of Figure~\ref{fig:prob} shows the spatial, metallicity and global velocity distribution of the four groups. 

The blue tail actually has spatial, metallicity and kinematic properties close to the majority group of red population, while the properties of the red tail is similar to the majority group of blue stars.  
Thus, the two populations identified by the probability cut have the properties representative of the true red/blue populations. This process does not bias more than weighting all the stars, moreover, it simplifies the separation of two populations. We adopt the hard cut on probability to separate the two populations and show that it works well in Section~\ref{SS:kin_mock}. 

\subsection{Model recovery}

\subsubsection{The recovery of the kinematics}
\label{SS:kin_mock}

To chemically tag stars, we could use the classic metallicity, or some proxy, such as color. We rely on relative values to perform the population separation.
The real metallicity distributions of the two populations of Sculptor have overlaps similar to the \emph{S025} samples. 
We illustrate the recovery of the spatial distribution, metallicity distribution and kinematics of the two populations using the results of the cored \emph{S025 1/3} and cusped \emph{S025 1/3} mock data. 

With mock data, we know the true kinematic properties of the two populations from their member stars. To assess how well our models are able to recover these true kinematic properties, we make two comparisons:
(1) we use the probabilities output by the model to identify red ($P_i^{red}>0.5$) and blue ($P_i^{blue}>0.5$) populations and then estimate the kinematic properties of each population from the data;
(2) we extract the model kinematic properties of the two populations using the best-fitting model parameters.

\begin{figure*}
\centering\includegraphics[width=\hsize]{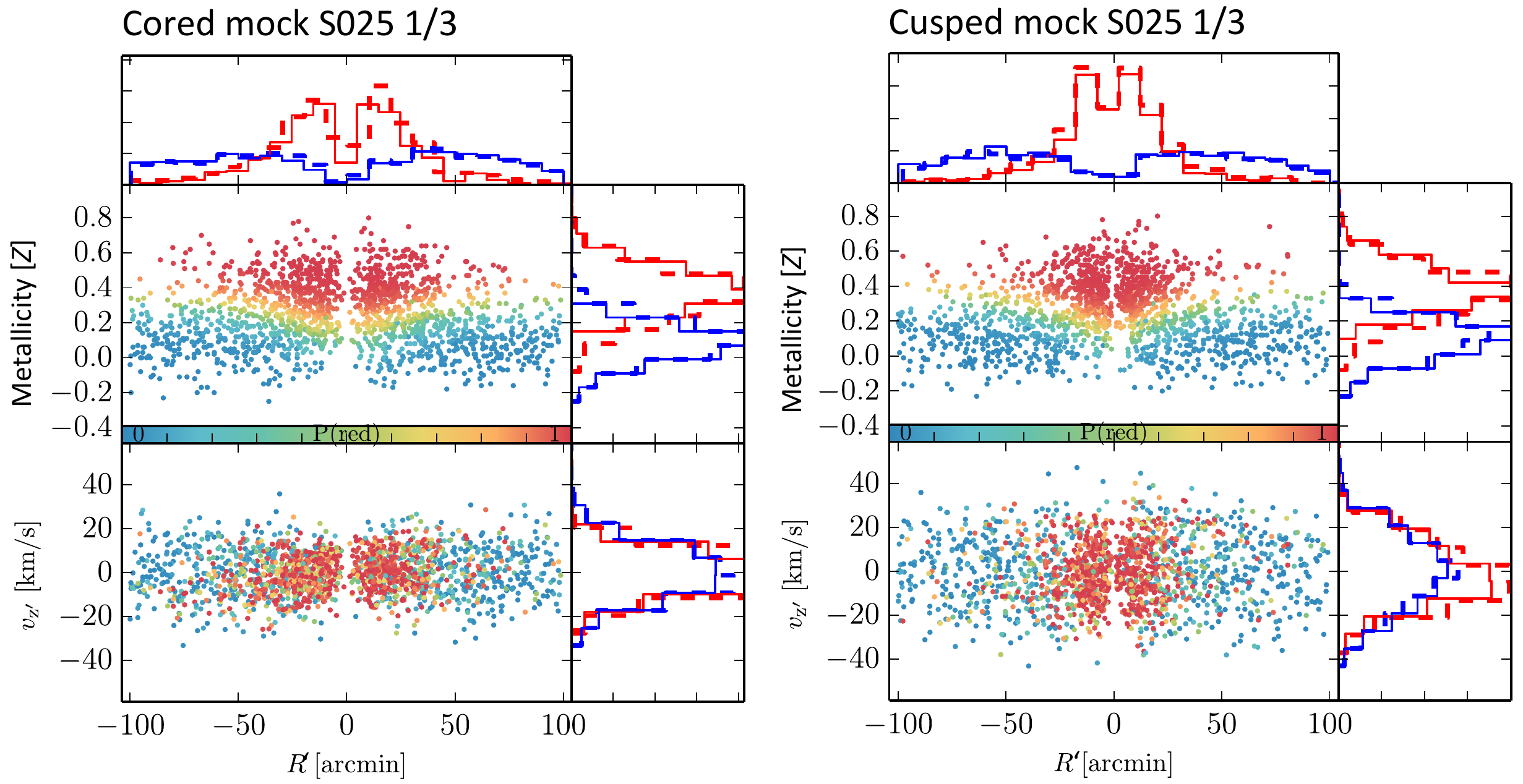}
\caption{The global properties of the two populations in the cored \emph{S025 1/3} (left) and cusped \emph{S025 1/3} (right) mock data sets and that recovered by our discrete chemo-dynamical model. There are five panels on each side. {\bf Top scatter panel}: The projected semi-major elliptical radius $R'$ vs. the metallicity $Z$.  {\bf Bottom scatter panel}: $R'$ vs. relative line-of-sight velocity $v_z$. Stars are plotted with points coloured by $P_i^\mathrm{'red}$ from blue (low) to red (high). The redder colors represent stars with higher probability to be in the red population, while the blue points represent stars with higher probability to be in the blue population. The red and blue dashed histograms 
show the true spatial, chemical and velocity distributions for the red and blue populations in the mock data. The solid histograms overlapped are the corresponding distributions of the red and blue stars identified by our chemo-dynamical models. Note that the velocity dispersion actually varies with radius, but here we plot the global velocity distribution for each population. 
}
\label{fig:chemdyn_mock}
\end{figure*}

Figure~\ref{fig:chemdyn_mock} shows the recovery of the global properties of the two populations. 
To represent the spatial distribution of the stars, we define the projected semi major elliptical radius $R'$:
\begin{equation}
\label{eqn:RR}
R' \equiv \rm{sign} (x')  \times \sqrt{x'^2 + (y'/\qbar')^2},
\end{equation} 
with $\qbar' = 0.90$ for the mock data. We use $R \equiv  \sqrt{x'^2 + (y'/\qbar')^2}$ representing the projected elliptical radius in the paper.

The scatter panel represents the distribution of stars in $R'$ vs. the metallicity $Z$, the bottom scatter panel represents $R'$ vs. relative line-of-sight velocity $v_z$. 
The true distributions (red and blue dashed histograms) of surface number density, metallicity and velocity of the two populations are generally recovered by the model-separated red and blue stars (red and blue solid histograms).

\begin{figure*}
\centering\includegraphics[width=\hsize]{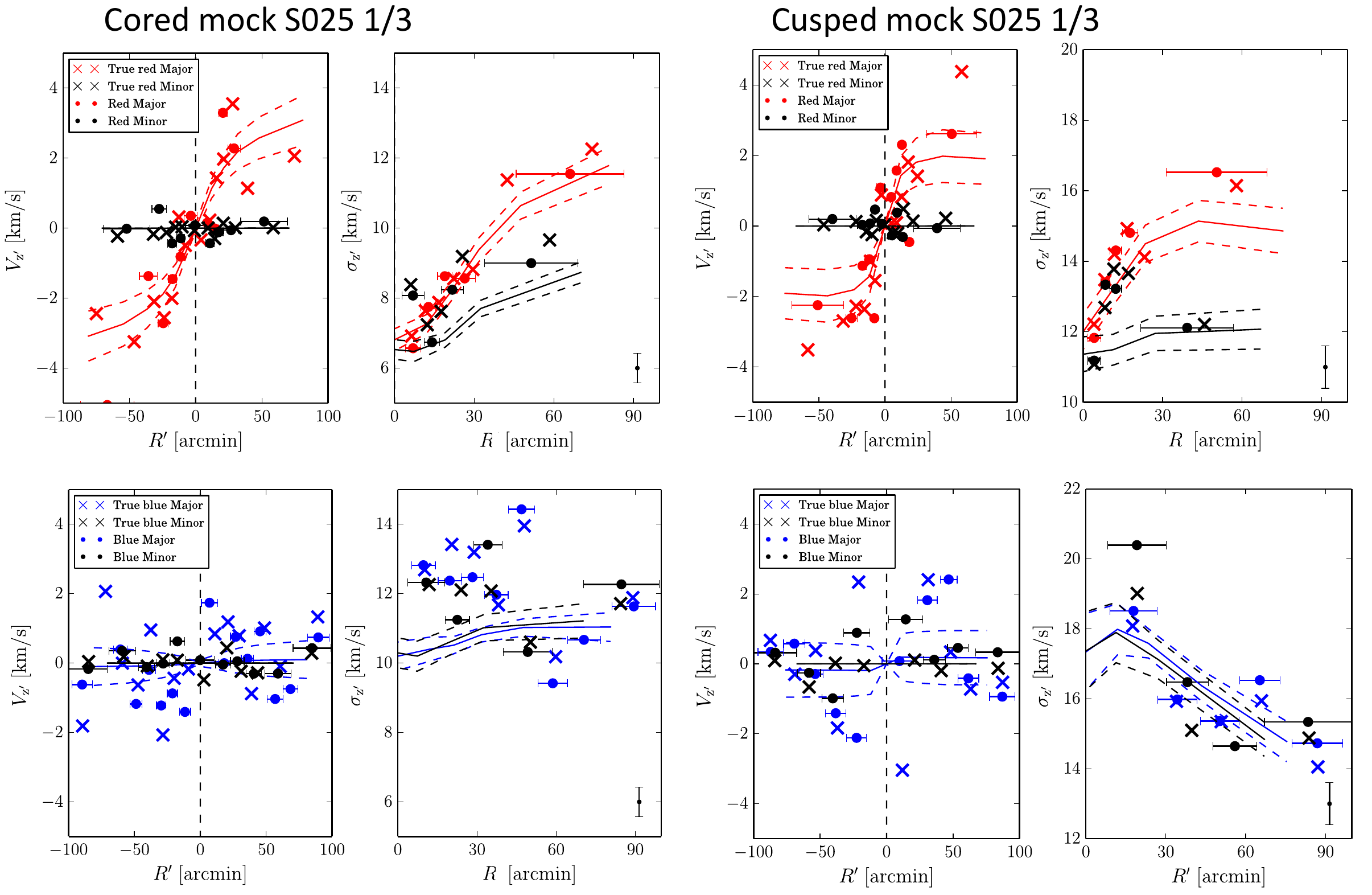}
\caption{The kinematic properties of the two populations in the cored \emph{S025 1/3} (left) and cusped \emph{S025 1/3} (right) mock data sets and that recovered by our discrete chemo-dynamical model. The four panels on each side show the mean velocity (left) and velocity dispersion (right) for the red population (top) and the blue population (bottom). 
In each panel, the colored symbols represent quantities along the major axis, and the black symbols represent that along the minor axis. 
The crosses represent the true value of velocity (velocity dispersion) binned from the true red (blue) stars, the corresponding values binned from red (blue) stars identified by our model are represented by dots, the corresponding model prediction with $1\sigma$ error is shown by the solid and dashed curves. The typical $1\sigma$ errorbar of the data is shown in the right-bottom corner.
The crosses and dots with the same color in each panel show the same trend as predicted by the model (solid line in the same color). In general, the JAM models describe the velocity dispersion profiles well, with the exception of the blue population in the cored mock data.}
\label{fig:kin_mock}
\end{figure*}

In Figure~\ref{fig:kin_mock}, we show the kinematic profiles of the two sets of mock data, and that recovered by the model.  
We extract the mean velocity and velocity dispersion profiles along the major and minor axis from the discrete data. The binning is performed along each axis with the stars in a cone of $45^o$ around that axis included. The mean velocity is binned along the projected semi-major/semi-minor axis, and we combine the stars in positive and negative directions together when calculating the velocity dispersion. Note that $R'$ represents the semi-major radius as defined in equation~\ref{eqn:RR} for the data binned along the major axis, while for the data binned along minor axis, $R'$ represent the semi-minor radius ( $\rm{sign} (y')  \times R $).

We use equal-population binning with 200 points in each bin; the bins do not overlap except for the outermost two bins.
The average radius of the 200 stars is taken as the value of radius of each bin, and the horizontal bar covers the radial range that the 200 stars span.
The model predictions for the red and blue populations are calculated from the kinematic maps for each population, with the same binning method as applied to the data.
The mean and scatter of the values are calculated with the models from every second of the last 50 steps of MCMC process.

For both mock datasets, we include a weak rotation in the red population with $\kappa^{red}_{true} \sim 0.3$. We see this rotation in the mean velocity profile of the true red populations (red crosses).  
The model matches this rotation with well-recovered $\kappa^{red}$, and the model-identified red stars follow the same trend (the red dots). There is no rotation in the true blue population, which is matched by the model, and followed by the model-identified blue stars. 

For the velocity dispersion profiles, our model-identified stars (dots) always follow the true red and blue populations (crosses). The JAM models describe the velocity anisotropy profiles well for the red populations of both mock datasets and the blue population in the cusped mock data. For an axisymmetric system following the assumptions of the JAM models, the information of velocity anisotropy is encoded in the difference of velocity dispersion along the major and minor axes. This difference is clearly seen in the velocity dispersion profiles binned from the true red and blue populations. In both cored and cusped mock data, the red population has a higher  
velocity dispersion along the major axis, which is matched well by the model predictions (the lines); the same is true for the blue population in the cusped mock dataset. 
However, the JAM model describes the kinematics less well for the blue population of the cored mock data, radially as well as azimuthally. This affects our estimates of the velocity anisotropy of the blue population and the mass profiles in the cored mock data as we show in Section~\ref{SS:mass_mock}.

\subsubsection{The rotation parametrs}
\label{SS:rot_mock}

\begin{figure}
\centering\includegraphics[width=\hsize]{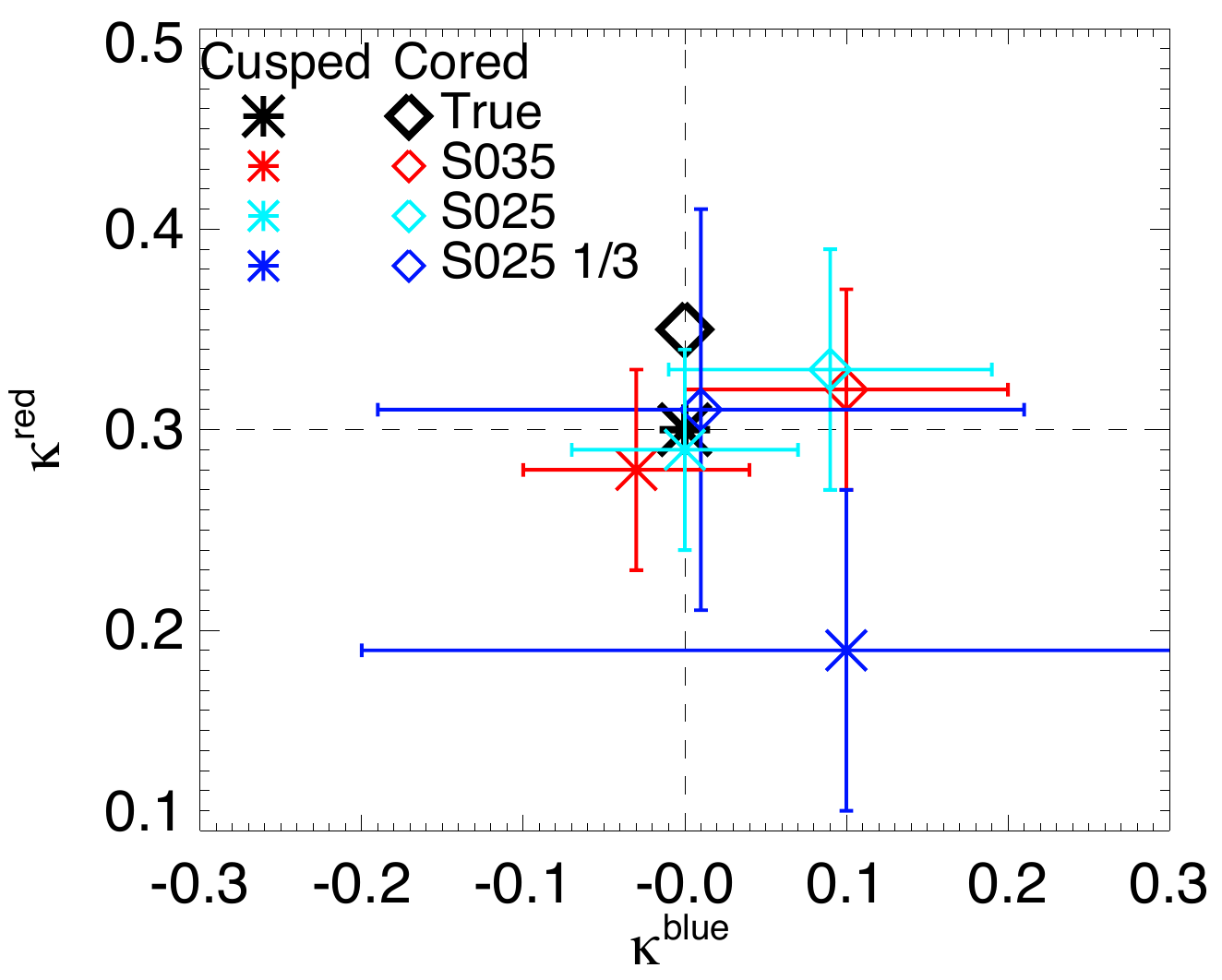}
\caption{The rotation parameters for the red ($\kappa^{red}$) and blue ($\kappa^{blue}$) populations recovered by models constrained by different mock data sets. The black diamond and asterisk are the true values for the cored and cusped mock data sets. The red,light blue and dark blue diamonds with errorbars are those recovered by the \emph{S035}, \emph{S025}, \emph{S025 1/3} cored samples. The asterisks represent the correspondingly values recovered by the cusped mock data sets. }
\label{fig:kpkr}
\end{figure}

The weak rotations are recovered well by our models as shown in Figure~\ref{fig:kpkr}. We find that:
\begin{itemize}
\item[-] The weak rotations in red populations and the zero rotations in the blue populations are generally recovered.

\item[-] The model tends to slightly underestimate the weak rotation of the red population. While a rotation parameter of $\sim 0.1$ for the blue population could be artificially introduced, and it tends to be positive (in the same direction as the red population).  

\item[-] The 1/3 samples have weaker ability to recover the rotation, with errorbars as twice large as that from the full samples.  
\end{itemize}

\subsubsection{The mass profiles and velocity anisotropies}
\label{SS:mass_mock}
The ability of our model to recover the mass profiles, especially the inner mass profiles depends critically on the ability to recover the kinematics of the two populations. 

\begin{figure}
\centering\includegraphics[width=\hsize]{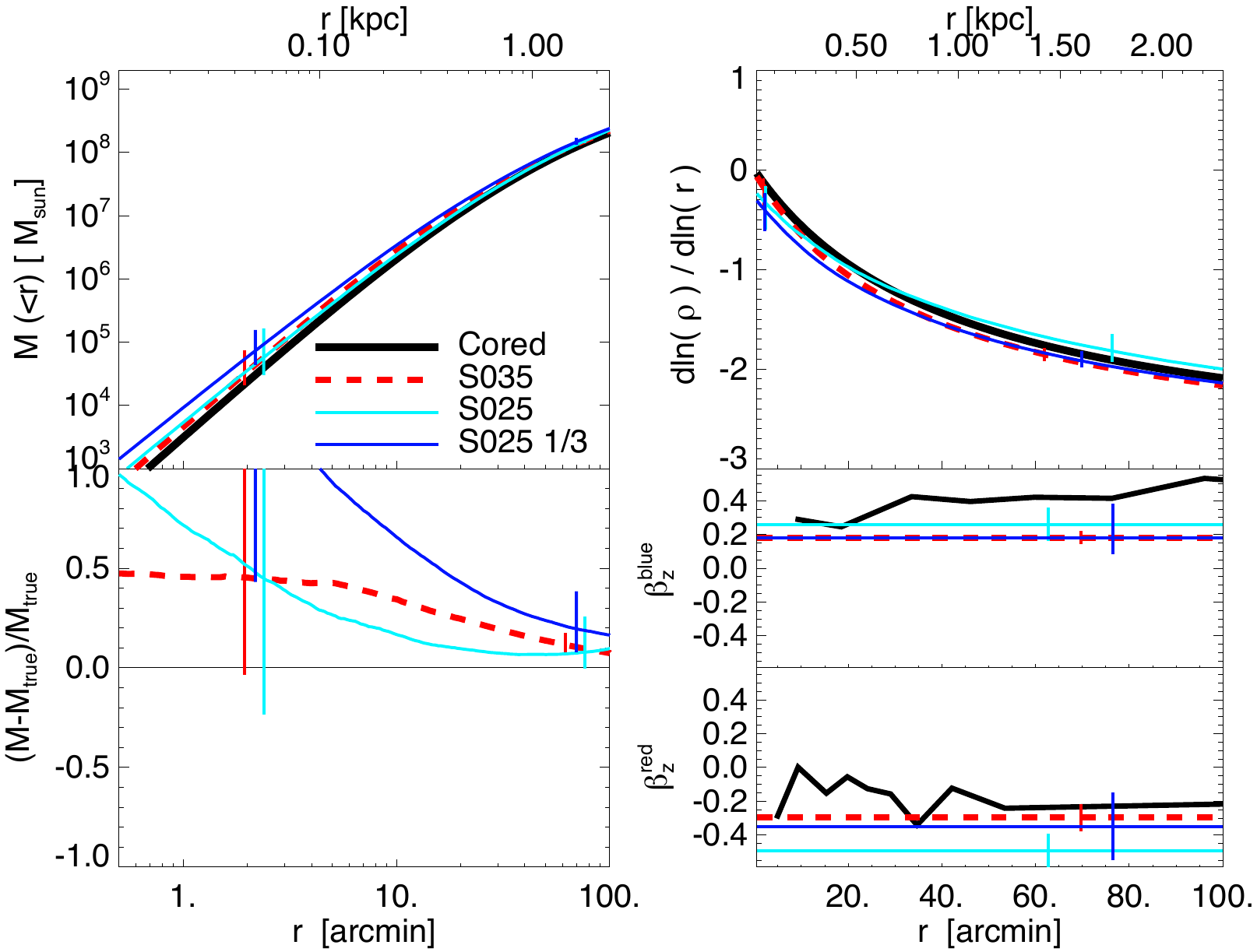}
\caption{The recovered mass profiles, density slope profiles, mass deviation from the true value, $\beta_z^{red}$ and $\beta_z^{blue}$ from models constrained by the three sets of cored mock data. In all five panels, the black line represent the true values. The red dashed, light blue and dark blue lines represent that recovered by the \emph{S035}, \emph{S025}, \emph{S025 1/3} mock data sets. 
Note that we have converted $\lambda_z$ to $\beta_z$ in the figure. The vertical short lines in all panels show the $1\sigma$ error at that particular position. }
\label{fig:core_mass}
\end{figure}

\begin{figure}
\centering\includegraphics[width=\hsize]{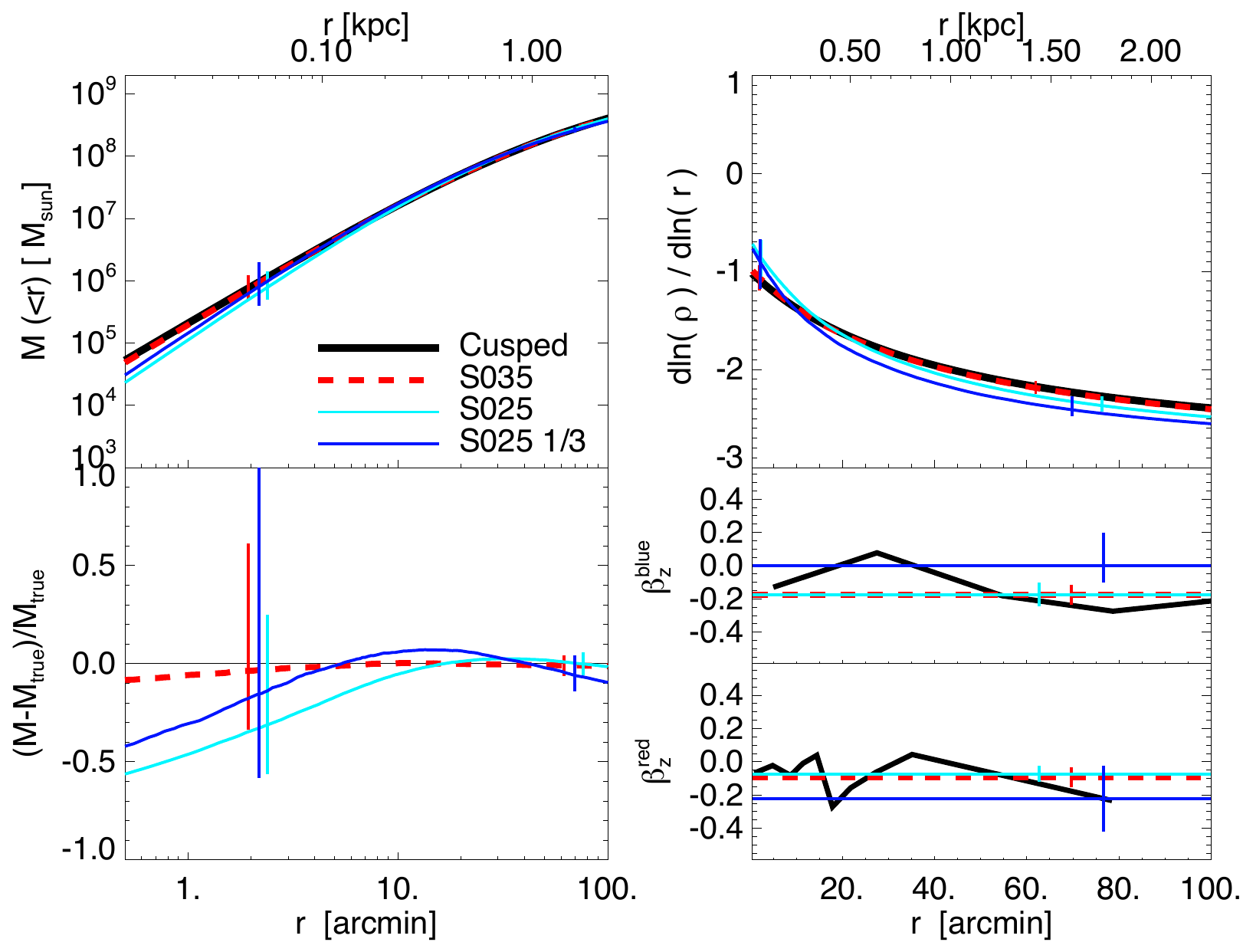}
\caption{The same as Figure~\ref{fig:core_mass}, but for the three sets of cusped mock data.} 
\label{fig:cusp_mass}
\end{figure}

Figure~\ref{fig:core_mass} and Figure~\ref{fig:cusp_mass} show the recovery of mass profiles, the deviation of mass profiles from the true mass, density slope profiles, and the velocity anisotropy parameter $\beta_z^{red}$ and $\beta_z^{blue}$ (converted from $\lambda_z^{red}$ and $\lambda_z^{blue}$) with different sets of mock data. In each panel the black line represents the true value.
The true velocity anisotropy profiles, varying with radius, are calculated from the mock data with full 6D information. 
We find that:

\begin{itemize}
\item[-] The mass profiles are generally well-recovered within $\sim 20\%$ uncertainties, except for larger uncertainties in the inner 10 arcmin ($\sim 200$ pc). And we generally find the medium value of the velocity anisotropy profile for each population by assuming constant velocity anisotropy parameter $\beta_z$ in our models. 

\item[-] The DM density slope profiles are fully recovered for the \emph{S035} samples. The uncertainties increase with decreasing metallicity separation of the two population and with decreasing number of data points. However, we can still distinguish the cored and cusped profiles with the \emph{S025 1/3} samples although with $1\sigma$ uncertainties of $\sim 0.3$ for the inner slope $\gamma$. 

\item[-] The recovery is generally worse for the cored mock data sets. The JAM models describe the velocity dispersions of the blue population less well (see also Fig~\ref{fig:kin_mock}). As a result, the velocity anisotropy parameters of the two populations (especially $\beta_{z}^{\mathrm{blue}}$), as well as the underlying gravitational potential, are recovered less well with the cored mock data. 
\end{itemize}

\section{Sculptor dwarf Spheroidal galaxy}
\label{S:Sculptor}
Sculptor is centered at $\alpha_{2000} = 1^{\mathrm{h}}0^{\mathrm{m}}6.36^{\mathrm{s}}$, $\delta_{2000} = -33^\circ42\arcmin12.6\arcsec$ with a position angle measured North through East of $85^\circ$, a systematic velocity $V_{\mathrm{sys}} = 110.6\pm0.5$\,\kms, and at a heliocentric distance $D = 79.3\,$kpc \citep{Battaglia2008a,deBoer2011}. The half light radius is $r_h = 0.26\pm0.039\,\mathrm{kpc}$ \citep{Walker2009jeans}.

\subsection{Data and models}
\subsubsection{Spectroscopic data}
\label{SS:dataspec}

We use the spectroscopic data from the Magellan/MMFS Survey of Stellar Velocities for Sculptor \citep{Walker2009Cat}. There are 1497 Red Giant Branch (RGB) stars with line-of-sight velocity and metallicity measurements in total in the sample. The lower limit of the tidal radius $r_t$ of Sculptor is $\sim 1300$ arcsec \citep{Walker2011}; we cut the data at 2000 arcsec $\sim 1.5 \, r_t$, leaving 1340 data points. Of these, we use only the 1218 good-quality stars for which both iron and magnesium indices and LOS velocities have been measured as the tracers for our chemo-dynamical models. 

We subtract the systemic velocity before creating the model. Since Sculptor has a large extent on the plane of the sky, its systemic motion may produce a non-negligible amount of perspective rotation \citep{Feast1961} which also has to be subtracted.  Expanding this perspective rotation in terms of the reciprocal of the distance $D$, and ignoring the negligible 
terms of order of $1/D^2$ or smaller, the following equation is obtained \citep{Glenn2006}:
\begin{equation}
     \label{eq:vzpr}
       v_{z'}^{pr} = 1.3790 \times 10^{-3} (x' \mu_{x'}^{sys}  + y' \mu_{y'}^{sys}) \,D \,\,$\kms$.
\end{equation}
For the global PM, we adopt $(\mu_{x'}^{sys} , \mu_{y'}^{sys} ) = (9 \pm 13, 2 \pm 13) \times 10^{-2}\,\, \mathrm{mas}\,\mathrm{yr}^{-1}$ as determined from \emph{HST} measurements \citep{Piatek2006}. 

We adopt the relative metallicity defined in \cite{Walker2009Cat},
\begin{equation}
	\label{eq:defrelZ}
	\Sigma Mg = -(0.079 \pm 0.002)(V- V_\mathrm{HB}) + \Sigma Mg',
\end{equation}
where $V - V_\mathrm{HB}$ is the offset in $V$-band luminosity from the Horizontal Branch (HB), with $V_\mathrm{HB} = 20.1$\,mag for Sculptor \citep{Walker2011}.  The slope quantifies the dependence of opacity on effective temperature and surface gravity, using luminosity as a proxy. The intercept, or reduced index $\Sigma Mg'$, represents the value of $\Sigma Mg$ that the star would have if it had the surface gravity and temperature of a HB star. Then taking the empirical calibration given by equation~(\ref{eq:defrelZ}) at face value, RGB stars of similar metallicity should have similar $\Sigma Mg'$, which, thus, will be used as the relative metallicity $Z$ of the stars.

The relative metallicity $\Sigma Mg'$ has not been calibrated with the absolute metallicity, so we do not use the metallicity distribution of the Milky Way halo from the literature in our background. Instead, we turn to the data. There are 19 stars outside 4000 arcsec with a roughly Gaussian velocity distribution centered at $-V_{sys}$, with no significant excess at the velocity of Sculptor. We consider these 19 stars as halo stars and obtain $\sigma^{bg} = 67$ \kms, $Z_{0}^\mathrm{bg} = 0.67$ and $\sigma_Z^\mathrm{bg} = 0.21$. The metallicity distribution of these 19 stars will be used as the fixed background parameters in our model. However our velocity dispersion $\sigma^{bg} = 67$ \kms is smaller than the halo's average velocity dispersion of 105 \kms \citet{XueXX2008}, so we take the value 105 \kms as the background velocity dispersion. 
\subsubsection{Surface number density}
\label{SS:datadens}

Our dynamical models require a surface number density profile for each tracer population in the form of an MGE. We have kinematic data for a sample of RGB stars that we will separate into a metal-rich (red) population and a metal-poor (blue) population in the model, so we require the number density profiles of both the red RGB stars and the blue RGB stars separately.

\begin{figure}
\centering\includegraphics[width=\hsize]{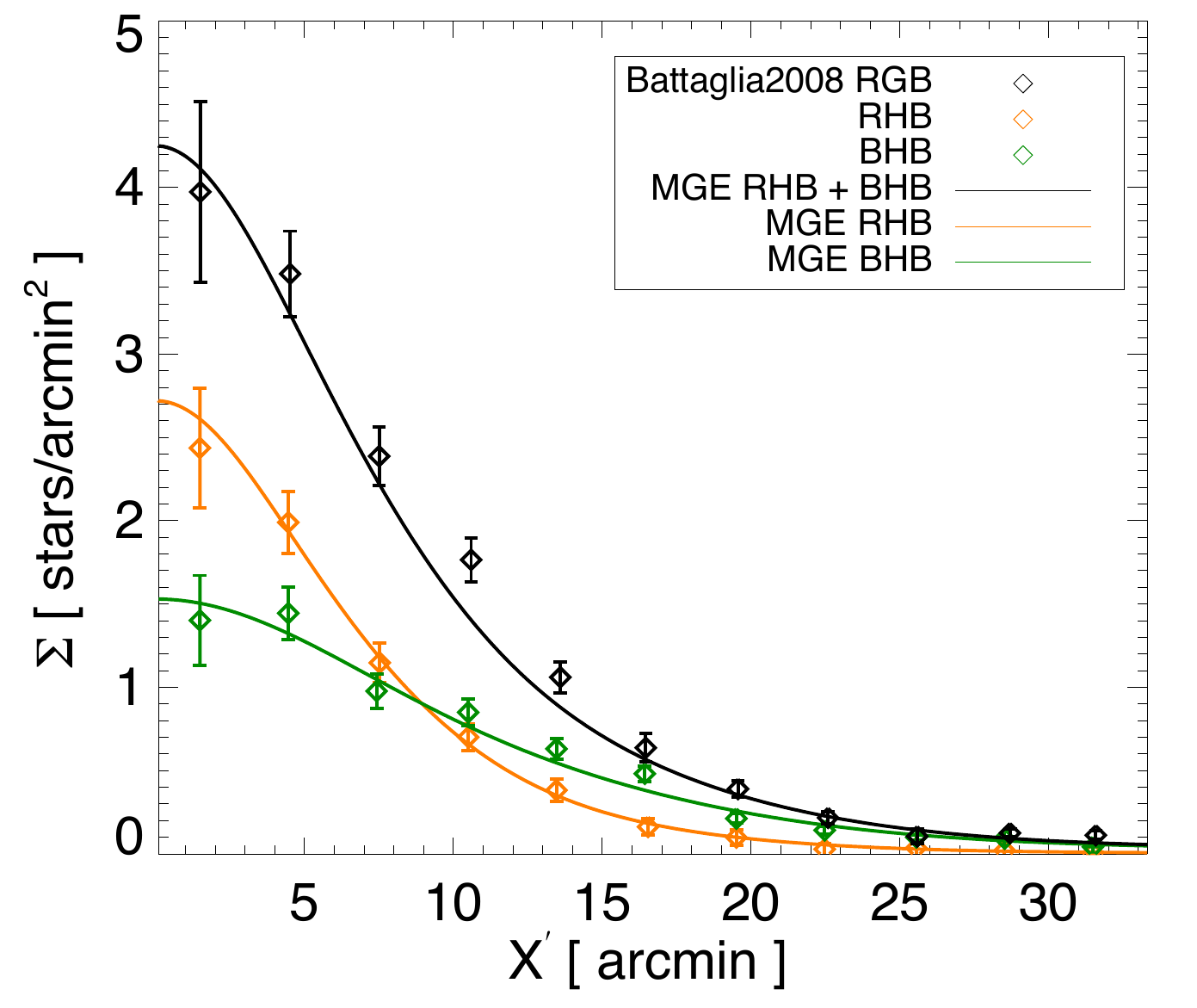}
\caption{Surface number density profiles for various type of stars. The orange and green diamonds with error bars are the RHB and BHB stars from \citet{Battaglia2008a}. The orange and green solid curves are the corresponding MGE fits. The black solid curve is the sum of the orange and green curves. The latter sum nicely matches the profiles of the RGB stars which is renormalized in scale. 
}
\label{fig:sculptor_numdens}
\end{figure}

\citet{Battaglia2008a} presented the surface number density profile for all RGB stars from ESO WFI photometry by counting the number of stars in elliptical shells with ellipticity of 0.28. The major axis profile is shown as black diamonds in Figure~\ref{fig:sculptor_numdens}. Separating the contributions of the red RGB stars and the blue RGB stars to this total RGB surface number density profile is difficult. In principle, we could leave the profiles of the red and blue populations completely free, with only their combined profile constrained by the observed total RGB surface number density profile. However, this will result in too many free parameters.

It is commonly assumed that the red and blue populations of RGB stars follow the number density profiles of red and blue horizontal branch (RHB and BHB) stars which can be clearly separated (e.g. \citealt{Amorisco2012Sculptor}; \citealt{Battaglia2008a}). \citet{Battaglia2008a} also constructed surface number density profiles of RHB and BHB stars; these are shown as orange and green diamonds respectively in Figure~\ref{fig:sculptor_numdens}. We fitted one-dimensional MGEs to these profiles; the ellipticity of the surface number density is measured to be approximately constant with radius, so we adopt the same projected flattening ${q'}_j = 0.72$ for each Gaussian $j$. The fits are shown as orange and green curves in Figure~\ref{fig:sculptor_numdens} and also listed in Table~\ref{tab:mge}. The total surface number density for the horizontal branch stars is the sum of these two profiles and is shown as the black line in Figure~\ref{fig:sculptor_numdens}.

\begin{table}
\caption{MGE fits of RHB and BHB surface number density profiles from \citet{Battaglia2008a}. $L_i$ is arbitrarily normalized. The left 3 Gaussians are the decomposition of the surface number density profile of RHB stars, while the right 3 are that of the BHB stars.  The combination of these six Gaussians will be treated as the total surface number density profile of RGB stars.}
\label{tab:mge}
\begin{tabular}{*{8}{l}}
\hline
\multicolumn{4}{l}{MGE fit RHB stars} & \multicolumn{4}{l}{MGE fit BHB stars} \\
\hline
 j &  $L_j$  &  $\sigma_j$ &  ${q'}_j$  & j &  $L_j$  &  $\sigma_j$ &  ${q'}_j$  \\
\hline
 1      &         0.50  &     193.0 &  0.72 &  4       &        0.52   &    346.9  &   0.72\\
 2      &         1.64  &     347.7 &   0.72 &  5       &        0.94   &    664.1  &  0.72\\
 3      &         0.57  &     602.1  &  0.72 &  6       &        0.07   &   1824.0 &    0.72\\
  \hline
 \end{tabular}
\end{table}


This combined RHB+BHB profile is in good agreement with the RGB profile so using the RHB and BHB profiles as proxies for the red and blue RGB profiles seems reasonable. However, instead of assuming that the red and blue RGB stars follow the number density profiles of RHB and BHB stars exactly, as previous studies have done, we allow the red and blue RGB profiles to be a linear combinations of the RHB and BHB profiles. Similar to the approach we used for the mock data, the resulting surface number density of the red population is then
\begin{equation}
	\label{eq:numdensred}
	\Sigma^\mathrm{red}(x',y') = h_1 \, \Sigma^\mathrm{RHB}(x',y') + h_2 \, \Sigma^\mathrm{BHB}(x',y'),
\end{equation}
and similarly for the blue population with fractions $1-h_1$ and $1-h_2$. As a result, the fractions $h_1$ and $h_2$ are two free \emph{density} parameters in addition to those 13 mentioned before in Section~\ref{SS:parameters}.

\subsubsection{Modelling steps}
\label{SS:steps}

In order to understand the ability of our model to distinguish between different DM halos for the real Sculptor, we first run a set of models with a generalised NFW halo with central density slope $\gamma$ free. Then two sets of models with different halo density slopes fixed, $\gamma = 0$ (cored halo) and $\gamma =1$ (cusped halo), are constructed to investigate the difference in the models caused by different DM central density slopes.
 
\begin{figure*}
\centering\includegraphics[width=\hsize]{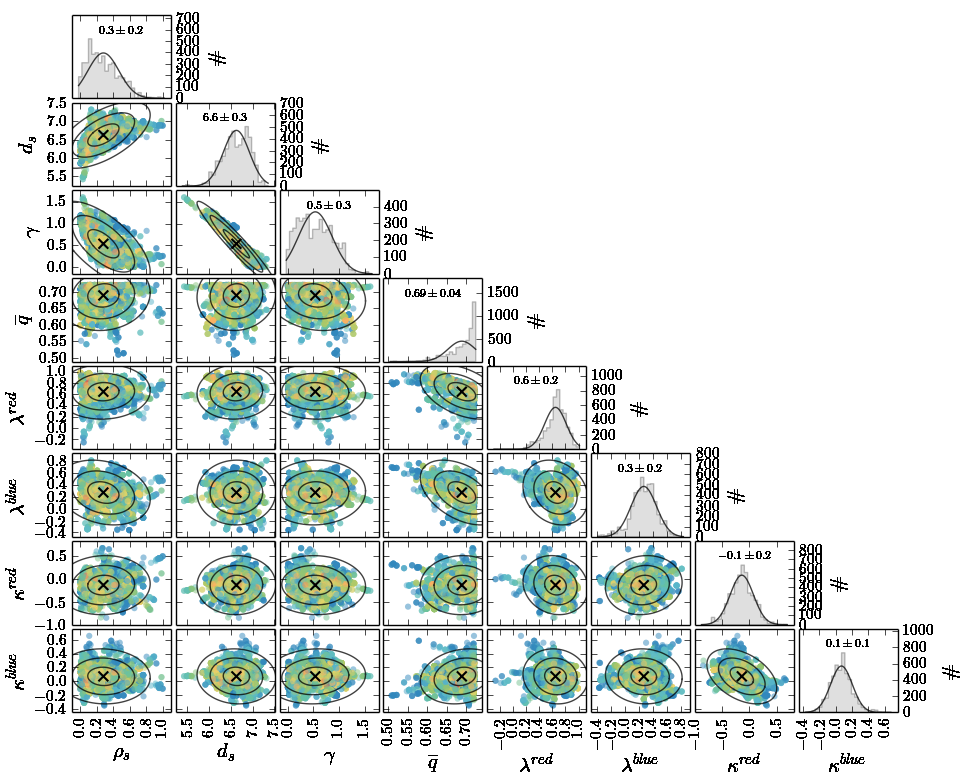}
\caption{MCMC post-burn distributions for our gNFW model of Sculptor. The scatter plots show the projected two-dimensional distributions, with the points coloured by their likelihoods from blue (low) to red (high). The ellipses represent the $1\sigma$, $2\sigma$ and $3\sigma$ regions of the projected covariance matrix. The histograms show the projected one-dimensional distributions. The parameters from left to right: DM scale density $\rho_s$, $d_s =\log( \rho_s^2 r_s^3)$ where $r_s$ is the scale radius, the inner density slope $\gamma$, the intrinsic flattening $\bar{q}$, velocity anisotropy parameter of the red population $\lambda^\mathrm{red}$ and the blue population $\lambda^\mathrm{blue}$, rotation parameter of the red population $\kappa^\mathrm{red}$ and the blue population $\kappa^\mathrm{blue}$. }
\label{fig:Smcmc_postburn}
\end{figure*}

We use the same MCMC process here as presented in Section~\ref{SS:mock_mcmc}. The MCMC post-burn distributions for the gNFW model are shown in Figure~\ref{fig:Smcmc_postburn}. We use the redefined $d_s = \log( \rho_s^2 r_s^3)$ parameter here to alleviate the degeneracy between potential parameters, however the inner density slope $\gamma$ is still degenerate with $d_s$. 

The best-fitting parameters obtained for each DM potential are presented in Table~\ref{tab:para}. Notice that, for convenience, $d_s$ has been converted to $r_s$, $\qbar$ has been converted to inclination angle $\vartheta$, and $\lambda^\mathrm{red}$ and $\lambda^\mathrm{blue}$ have been converted to $\beta_z^\mathrm{red}$ and $\beta_z^\mathrm{blue}$. 
In the generalised NFW model, the DM density slope $\gamma$ is degenerate with $d_s$, which corresponds to the degeneracy between $\gamma$ and the DM scale radius $r_s$, thus there is large uncertainty in $\gamma$. 
We will refer to these parameters in the following sections when discussing the main results.

\begin{table*}
\caption{The best-fitting parameters obtained by the MCMC process for three set of models with different  DM halos: generalised NFW halo with central density slope $\gamma$ free, cored halo with $\gamma = 0$ and cusped halo with $\gamma = 1$. The parameters are presented in two rows for each model. First row from left to right: DM scale density $\rho_s$, DM scale radius $r_s$, central density slope $\gamma$, the inclination angle $\vartheta$,  the background fraction $\epsilon$, the fraction $h_1$ and $h_2$ of the RHB and BHB surface number density profiles contributing to the red population, and the maximum likelihood $L_{max}$.
Second row from left to right: mean metallicity $Z_0^\mathrm{red}$ and metallicity spread $\sigma_Z^\mathrm{red}$, velocity anisotropy in the meridional plane $\beta^\mathrm{red}_z$  and the rotation parameter $\kappa^{red}$ of the red population, as well as the corresponding parameters for the blue population. Note that for convenience the recast parameter $d_s$ has been converted to $r_s$, $\qbar$ has been converted to inclination angle and $\lambda^\mathrm{red}$ and $\lambda^\mathrm{blue}$ have been converted to $\beta_z^\mathrm{red}$ and $\beta_z^\mathrm{blue}$.}
\label{tab:para}
\small
\scriptsize
\footnotesize
\begin{tabular}{llllllllll}
\hline
\hline
 DM & $\rho_s [M_{\sun} /\mathrm{pc}^3]$ & $r_s  [\mathrm{pc}]$ & $\gamma$  & inclination  & $\epsilon$ $\%$ & $h_{1}$ & $h_{2}$   & $L_{max}$&\\
\\
   & $Z_0^\mathrm{red}$  & $\sigma_Z^\mathrm{red}$  & $\beta_z^\mathrm{red}$ &  $\kappa ^{red}$&  $Z_0^\mathrm{blue}$  & $\sigma_Z^\mathrm{blue}$   &  $\beta_z^\mathrm{blue}$  & $\kappa ^{blue}$& \\
  \hline 
  \hline
 gNFW & $0.3\pm0.2$  &  $350^{+300}_{-100}$ & $0.5\pm0.3$ & $75^{+14}_{-7}$ &$0.8\pm0.1$  &$0.9\pm0.1$ &  $0.1\pm0.1 $ & -23078 &\\ 
 \\
   & $0.37\pm0.01$  & $0.079\pm0.006$  &   $0.44^{+0.1}_{-0.12}$ & $-0.1\pm0.2$&  $0.282\pm0.007$ & $0.047\pm0.006$   &$0.26^{+0.10}_{-0.13}$& $0.1\pm0.1$ & \\
  \hline
  cored & $0.5\pm0.2$  & $370^{+100}_{-70}$   & 0  & $79^{+10}_{-5}$  & $0.8\pm0.1$  & $0.95\pm0.05$ & $0.1\pm0.1$ & -23079 &  \\
 \\
   &$0.37\pm0.01$ & $0.07\pm0.01$  & $0.47^{+0.09}_{-0.11}$ & $-0.2\pm0.3$ & $0.281\pm0.006$ & $0.049\pm0.005$   & $0.36^{+0.11}_{-0.14}$ & $0.1\pm0.2$\\
 
 \hline
 cusped & $0.08\pm0.07$ & $570^{+700}_{-150}$ & 1 & $72^{+16}_{-8}$  & $0.8\pm0.1$  &$0.93\pm0.06$ &$0.1\pm0.1$  & -23079& \\
 \\
   &  $0.37\pm0.01$ & $0.08\pm0.01$ &  $0.48^{+0.09}_{-0.11}$ & $-0.2\pm0.2$ &  $0.28\pm0.01$ & $0.047\pm0.005$  & $0.25^{+0.13}_{-0.16}$  & $0.1\pm0.1$  \\

  \hline
  \hline
 \end{tabular}
\end{table*}

\subsection{Results for Sculptor}
\label{S:results}

\subsubsection{Two-population spatial, chemical and velocity distributions}
\label{SS:resdistr}
Following our treatment of the mock data, in the model for which the best parameters were obtained, the stars can be separated via probability as calculated by equation~(\ref{eq:Pik}). The stars with $P_i^\mathrm{'red} > 0.5$ ($P_i^\mathrm{'blue} > 0.5$) will be treated as red (blue) stars,  while the stars with $P_i^\mathrm{'bg} > 0.5$ are contaminant stars.

The models with different DM halos predict different kinematics for each population. For a single star $i$, its probabilities $P_i^\mathrm{'red}$  and  $P_i^\mathrm{'blue}$ are different from model to model, so the group of red and blue stars are different from model to model.  The best-fitting model with a generalised NFW halo identifies 445 red stars and 653 blue stars,
the best-fitting model with a cored halo identifies 444 red stars and 646 blue stars and the best-fit model with a cusped halo identifies 376 red stars and 696 blue stars.  

Excluding $\sim 70$ stars being selected as contaminant stars for each model, there are another $\sim 50$ stars for which neither the blue or red probabilities are larger than 0.5 and so are excluded in what follows.

%

\begin{figure}
\centering\includegraphics[width=\hsize]{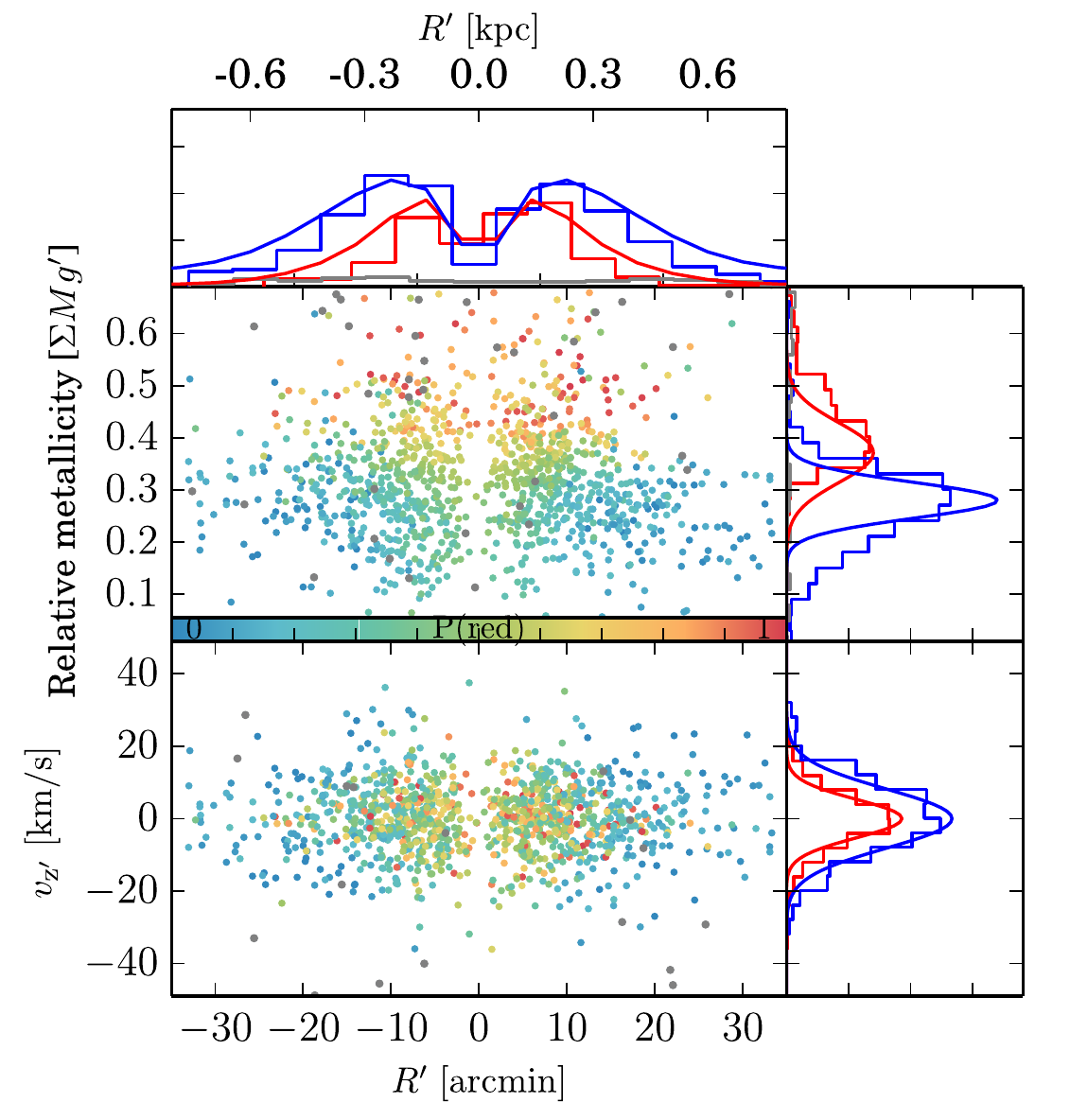}
\caption{{\bf Top scatter panel}: The projected semi-major elliptical radius $R'$ vs. the relative metallicity $\Sigma Mg'$.  {\bf Bottom scatter panel}: $R'$ vs. relative line-of-sight velocity $v_z$. The grey points represent the contaminant stars. Stars belonging to Sculptor are plotted with points coloured by $P_i^\mathrm{'red}$ from blue (low) to red (high).  The redder colors represent stars with higher probability to be in the red population, while the blue points represent stars with higher probability to be in the blue population. The red and blue histograms show the spatial, chemical and velocity distributions for the stars identified as red and blue, while the grey histograms are for the contaminant stars. The solid curves over-plotted are the model predictions for each population. The model-predicted number of stars per radial bin are inferred from their surface number density profiles. Note that the velocity dispersion actually varies with radius, while here we plot the global velocity distribution for each population.}
\label{fig:sculptor_discrdistr}
\end{figure}

Figure~\ref{fig:sculptor_discrdistr} shows the separation of the stars in the best-fitting cored model. 
As in Figure~\ref{fig:chemdyn_mock}, 
$R'$ is the projected semi major elliptical radius but with $\qbar' = 0.72$ for Sculptor. 

The red and blue histograms are directly constructed with the model-identified red and blue stars ($P_i^\mathrm{'red} > 0.5$ for red and $P_i^\mathrm{'blue} > 0.5$ for blue), while the grey histograms are for the contaminant stars. 
The solid curves on the histograms are the model predictions for each population. Because the MCMC chains have ``memory'' of the previous step, consecutive steps are not independent, thus all the model curves are constructed with every two steps of the last 50 steps of the MCMC process.   

We obtained surface number density fractions $h_1\sim 1$ and $h_2 \sim 0$,
indicating that the model surface number density profile of the red population is thus very close to that of the RHB stars, which are more concentrated, while that of the blue population is dominated by the shape of BHB stars, which are more extended.  This assumption in previous two-component dynamical models for Sculptor (e.g. \citealt{Amorisco2012Sculptor}; \citealt{Battaglia2008a}), our model thus shows is reasonable. 
The radial distributions of the model-identified stars are consistent with the model predictions $dN(R')$--inferred from the surface number density profiles $\Sigma(R)$, with $dN(R') = dN(R) /2 = \pi R \Sigma(R) dR$--but not exactly the same, because the stars with discrete velocity measurements have selection functions that vary with radius.
The metallicity distributions of the two populations show significant overlap but are clearly distinguishable. The red population has a higher metallicity spread than the blue population.

The $\sim 70$ stars classified as contaminant stars are shown in grey in Figure~\ref{fig:sculptor_discrdistr}. The background stars are selected out by the model naturally. They are generally uniformly distributed in radius, and they have a wide metallicity distribution ($Z_0^{bg} = 0.57$, $\sigma_Z^{bg} = 0.31$) and a wide velocity distribution ($\sigma_0^{bg}$ = 111 \kms). These properties are generally consistent with the input background parameters. 

\subsubsection{Two-population kinematics}
\label{SS:reskin}
\begin{figure*}
\centering\includegraphics[width=\hsize]{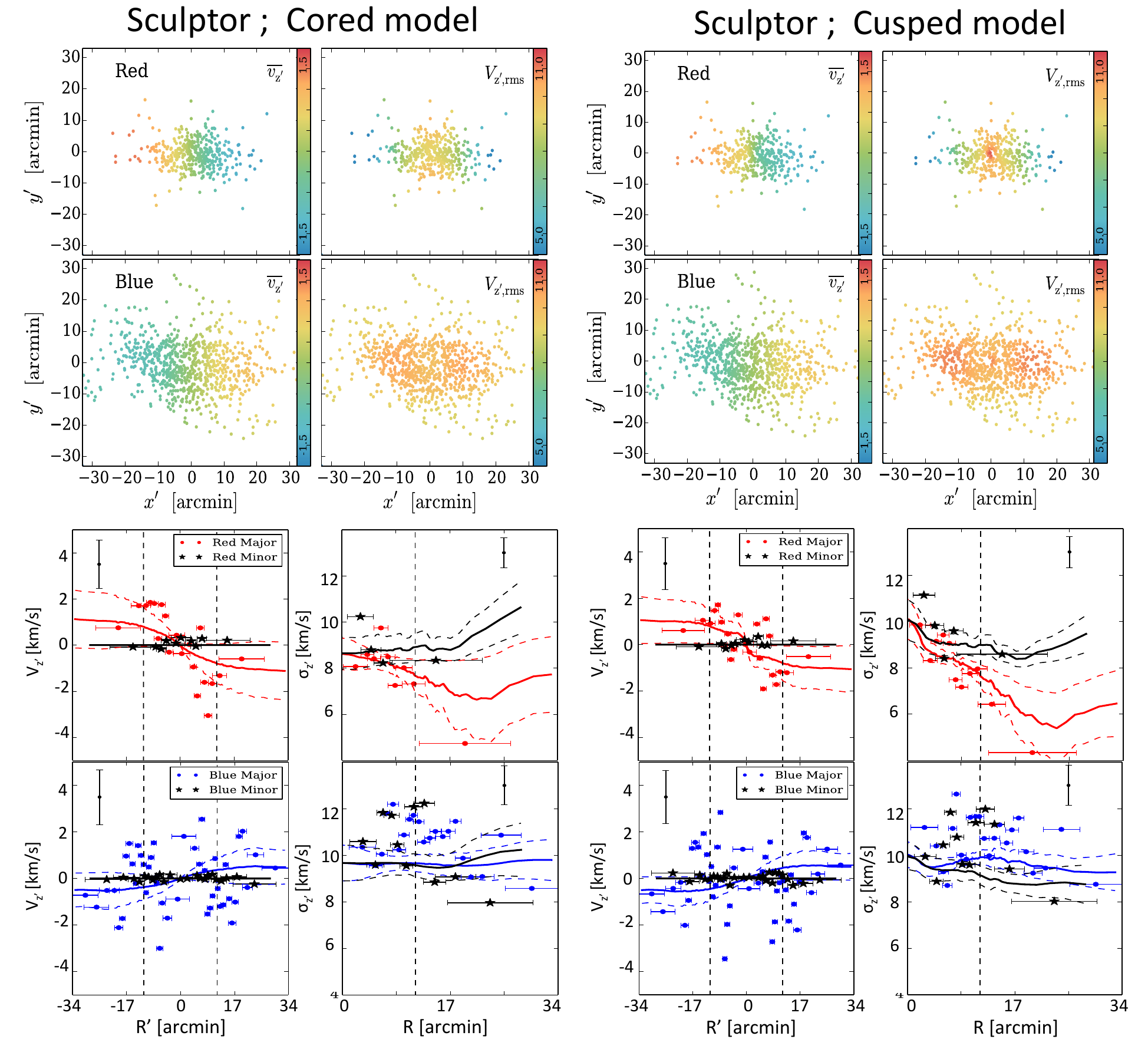}
\caption{Line-of-sight kinematics of the two populations in Sculptor, adopting a cored (left) or cusped (right) DM halo. {\bf Upper panels}: model-predicted mean velocity and velocity dispersion maps for the red (top) and blue (bottom) populations. Each point represents a star position coloured with the velocity or velocity dispersion values, which are scaled as indicated by the corresponding colour bars.
{\bf Lower panels}: comparison of data and model mean velocity and velocity dispersion profiles for the red (top) and blue (bottom) populations. The coloured (black) line represents the profile along the major (minor) axis predicted by the best-fitting model, while the coloured dots (black stars) with error bars represent the data along the major (minor) axis after spatial binning of the model-identified red or blue stars.}
\label{fig:kin_freekap}
\end{figure*}

The kinematics of the red and blue populations predicted by the best-fitting models with cored and cusped DM halos are shown in Figure~\ref{fig:kin_freekap}. 
The best-fitting model with a generalised NFW halo is in between that of a cored and a cusped halo. 
For each model, the upper panels show the model-predicted mean velocity and velocity dispersion maps for the red (top) and blue (bottom) populations. Each point represents a star position coloured with the corresponding  velocity and velocity dispersion values.  

We extract the mean velocity and velocity dispersion profiles to see how well the model matches the data.
We bin the data along the major and minor axis, as we did in Section~\ref{SS:kin_mock}.  $R'$ is the projected semi-major/semi-minor elliptical radius for the data binned along major/minor axis, and $R$ is the projected elliptical radius.  
Equally-populated radial bins are used. Due to the limited number of stars, here we put 80 points in each bin with 40 points of overlap in bins close to each other. 
The lower panels of Figure~\ref{fig:kin_freekap} shows the binned mean velocity and velocity dispersion profiles for the red (top) and blue (bottom) populations. 
The model-identified red and blue stars show distinct velocity dispersion profiles.

The red stars generally have a lower velocity dispersion than the blue stars. In addition, the red stars have a lower dispersion along the major axis than the minor axis (the red dots are lower than the black star symbols), while the blue stars have a higher dispersion along the major axis (the blue dots are higher than the black star symbols). 
Although the cored model and the cusped model identified the red and blue stars independently, the kinematic properties of each population identified by the two different models are consistent with each other. 

We note that the cored DM halo models predict flat velocity dispersion profiles for both populations, and the dispersion only significantly decreases with radius for the red population along the major axis. The cusped DM halo models always predict central peaks for the velocity dispersions of both populations, the decline of the velocity dispersion profiles with radius is more obvious. 
The anisotropy of the red population is matched equally well by the cored and cusped model, while the cusped model matches the anisotropy of the blue population better than that of the cored model. 
However, even though the model predictions from the cored and cusped models are different, with the limited data points, we do not have a statistically-significant preference for either model. The maximum likelihood of the cored and cusped models are equally good.

A $1\sigma$ significant internal rotation is revealed in the red stars as matched by the models.

\subsubsection{The mass profiles}
\label{SS:rescorecusp}
\begin{figure*}
\centering\includegraphics[width=\hsize]{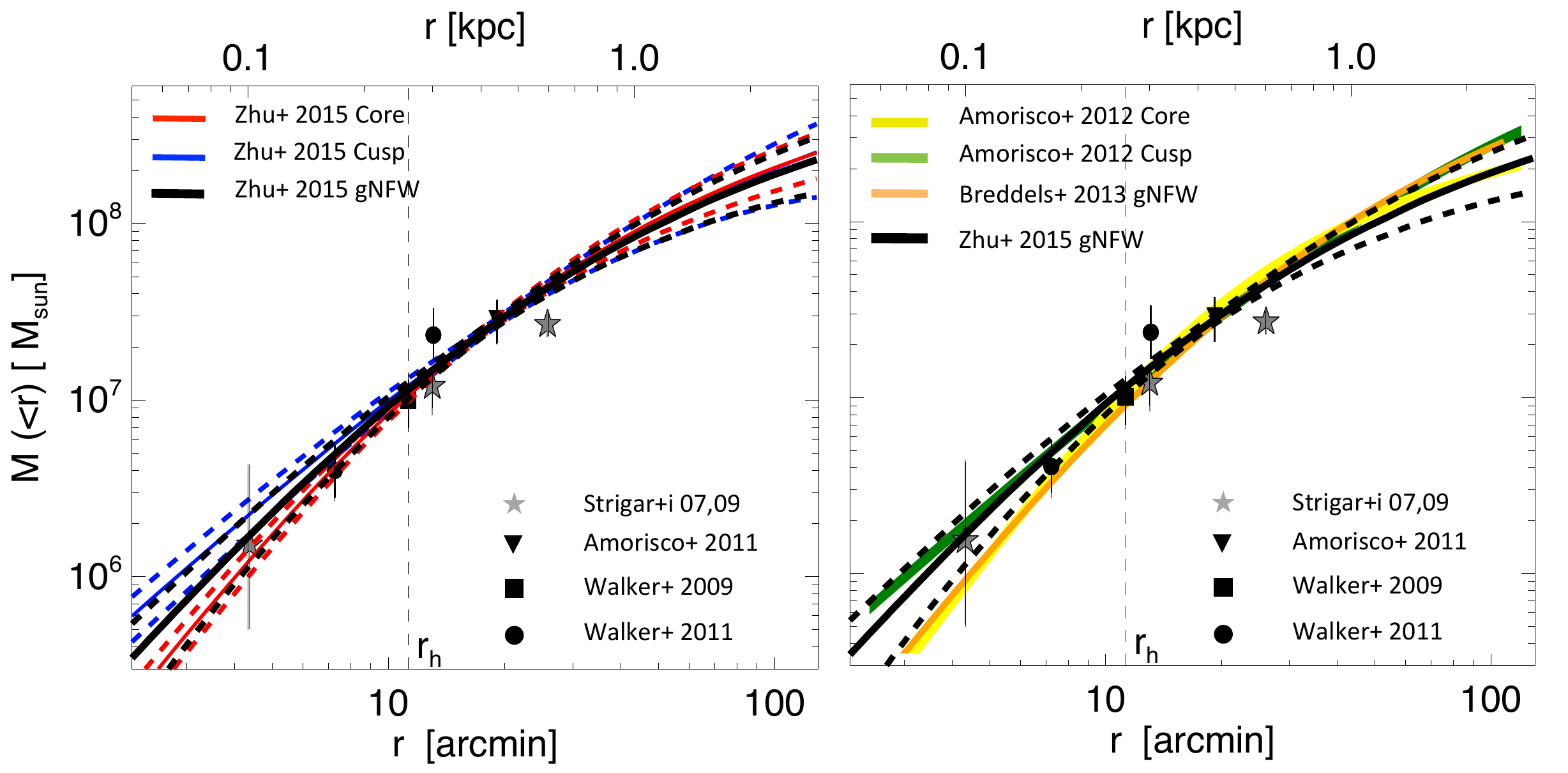}
\caption{Total enclosed mass profiles for Sculptor. {\bf Left panel}: the black solid and dashed curves are the mass profile for the generalised NFW halo with $1\,\sigma$ uncertainty, red curves are those for the cored DM halo and the blue curves are those for the cusped DM halo obtained by our discrete axisymmetric chemo-dynamical models. The symbols represent the virial masses obtained at different radii from different papers as labeled. The dashed vertical line indicates the position of the half-light radius $r_h$.
{\bf Right panel}: again the black solid and dashed curves are our mass profiles for the generalised NFW halo, the yellow and green thick curves are the mass profiles of the cored and cusped halos respectively obtained by a two-component spherical Michie-King model \citep{Amorisco2012Sculptor}, the orange line is obtained by a spherical Schwarzschild model with a generalised NFW DM halo \citep{Breddels2013}. }
\label{fig:sculptor_enclmass}
\end{figure*}

The best mass profiles obtained by different DM halo models are shown in Figure~\ref{fig:sculptor_enclmass} and compared with previous estimates. 
In the left panel, the black solid and dashed curves are the mass profiles of the generalised NFW halo model with $1\,\sigma$ uncertainty, the red curves are those of the cored DM halo model and the blue curves are those of the cusped DM halo model. 
The largest difference between the mass profiles of the cored and cusped halos is seen in the inner 200 pc ($\sim 8.7$ arcmin), where the generalised NFW model has a large uncertainty with the central density slope converging to $\gamma = 0.5\pm0.3$.
Hence, the cored and cusped halos are still not distinguishable statistically with the present data.

The symbols represent the virial mass estimates obtained at different radii in previous studies by (\citealt{Strigari2007}; \citealt{Strigari2008}; \citealt{Walker2009jeans};  \citealt{Walker2011}; \citealt{Amorisco2011})\footnote{For \citet{Walker2011}) we use the value in their Figure 10, not the value in the table}.
Most of the virial mass estimates are consistent with our mass profiles, and do not distinguish between a cored or a cusped DM profile, except for \citet{Walker2011}. The latter authors used the two populations in Sculptor and obtained the mass at the half-light radius of the metal-rich and metal-poor populations independently and claimed that the cusped DM halo can be statistically excluded when Sculptor is assumed to be spherical. However, when an elliptical radius instead of spherical radius is used, \citet{Walker2011} obtained $\gamma = 0.6^{+0.26}_{-0.32}$ for Sculptor, which is still consistent with our results. 
 
Under the assumption of sphericity, the lower limit of the slope obtained by \citet{Walker2011} is consistent with the mass profile of the cored and the generalised NFW DM halo models we obtained, but their virial mass estimate for the metal-poor population is higher than the others. 
Different dynamical assumptions will affect the separation of the two populations of stars, so as a result, \citet{Walker2011} assigned fewer stars to the metal-poor population than our model.
 The velocity dispersion obtained and used by \citet{Walker2011} to calculate the virial masses were $\sigma_{0}^{red} = 6.5^{+0.4}_{-0.5}$ km s$^{-1}$ and  $\sigma_{0}^{blue} = 11.6^{+0.6}_{-0.6}$ km s$^{-1}$. With the two population of stars separated by our model in Section~\ref{SS:resdistr}, the mean velocity dispersion is $\sigma_{0}^{red} = 7.4\pm0.5$ km s$^{-1}$ and $\sigma_{0}^{blue} = 10.6 \pm 0.5 $ km s$^{-1}$, these values decrease the slope of the mass profile to match the mass profile obtained by our model.  

In the right panel of Figure~\ref{fig:sculptor_enclmass}, our mass profile for a generalised NFW halo is plotted, with the mass profiles from \citet{Amorisco2012Sculptor} for cored and cusped halos (yellow and green curves). They used a two-component Michie-King phase-space model to fit the dispersion profiles of the metal-rich and metal-poor populations simultaneously. Their mass profiles for cored and cusped DM halos both match our corresponding estimates. The cored DM halo is preferred in their model although the cusped DM halo is not excluded. Finally, the orange line represents the mass profile obtained by a spherical Schwarzschild model with a generalised NFW halo \citep{Breddels2013}, consistent with our estimates at $1\sigma$ confidence. 

\section{Discussion}
\label{S:discussion}
\subsection{Velocity anisotropy}
\label{SS:discvelani}

\begin{figure}
\centering\includegraphics[width=\hsize]{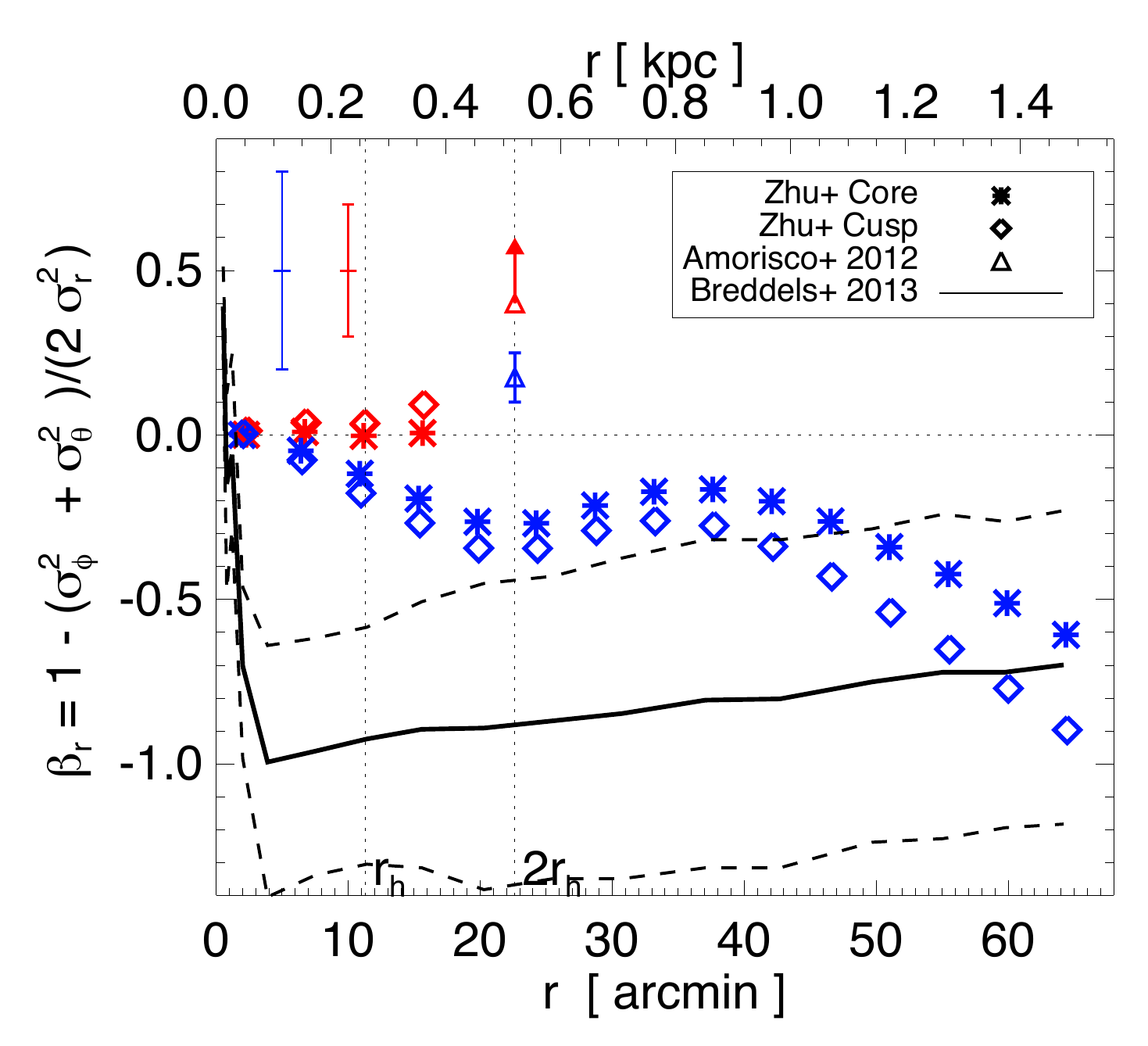}
\caption{The velocity anisotropy $\beta_r$ profiles as function of the intrinsic radius $r$.  Red symbols represent that for the metal-rich and blue the metal-poor population. The asterisks and diamonds are for the cored model and cusped model respectively, with the typical error bar shown in the left-top corner.
The red and blue triangle are the velocity anisotropy of the red and blue population given by the two-component model of \citet{Amorisco2012Sculptor}. The black solid and dashed lines are the velocity anisotropy from the single-component Schwarzschild model with $1\sigma$ error \citep{Breddels2013}. The two vertical dashed lines indicate 1 and 2 half light radii.
}
\label{fig:betar}
\end{figure}

It is still debated whether Sculptor is radially anisotropic or tangentially anisotropic (e.g., \citealt{Breddels2013}; \citealt{Amorisco2012Sculptor}).  
In a spherical model,  
\begin{equation}
\beta_r = 1-\frac{ \sigma_{\phi}^2 +  \sigma_{\theta}^2 }{2 \sigma_{r}^2 },
\end{equation}
is used to describe the velocity anisotropy of the system, where $\sigma^2_{\phi} = \overline{v_{\phi}^2} - \bar{v}_{\phi}^2 $, $\sigma^2_{r} = \overline{v_r^2}$ and $\sigma^2_{\theta} = \overline{v_{\theta}^2}$. In order to compare with previous results, we transform the second velocity moments from cylindrical polar coordinates to spherical coordinates and infer $\beta_r$ from our axisymmetric models. In Figure~\ref{fig:betar}, the velocity anisotropy profiles are calculated using the models within the $1\,\sigma$ confidence level from every second step of the last 50 steps of the MCMC process; the error bars indicate the typical spread among these models. The red and blue symbols represent the red and the blue populations. The asterisks and diamonds are for the cored and cusped models, respectively. 

We find that the red population is nearly isotropic, while the blue population is close to isotropic at small radius and becomes mildly tangentially anisotropic outwards. Also the blue population shows a higher degree of tangential anisotropy in the cusped model than in the cored model.
As shown in Figure~\ref{fig:kin_freekap}, the cusped model matches the anisotropy of the blue stars better, thus we prefer the higher tangential velocity anisotropy of the blue population from the cusped model. 

\citet{Battaglia2008a} separated the two populations using hard cuts on metallicity, and constructed a Jeans model for each population independently. They found both populations to be consistent with radial anisotropy.
\cite{Amorisco2012Sculptor} used the data from \cite{Battaglia2008a}, and created a two-component Michie-King phase-space model. They found that in a cored DM halo model, a nearly isotropic metal-poor stellar population is preferred (${\beta_r} \approx 0.1$), whilst a cusped DM halo model favours a mild radial velocity bias (${\beta_r} \approx 0.25$). The metal-rich population requires a high degree of radial anisotropy (${\beta_r} > 0.4$), irrespective of the choice of DM halo.  Their results may be biased because no tangential velocity anisotropy is allowed in Michie-King DFs. Even so, the metal-rich population being more radial anisotropic than the metal-poor population is consistent with our findings.

\citet{Breddels2013} created single-component orbit-based Schwarzschild models of Sculptor.
They binned the data in radius assuming a single population, and showed that the LOS velocity distribution of Sculptor deviates from a Gaussian distribution: the velocity distributions for the bins at small radii have a narrow peak, while the velocity distributions are more flat-topped for the bins in the outskirts.
They found mild radial anisotropy in the inner 0.1 kpc that dropped to highly tangential velocity anisotropy with $\beta_r \sim - 1.0$ at all radii outside 0.2 kpc. 
However, our models indicate that Sculptor clearly has two populations with different spatial and velocity distributions. The red population is more spatially concentrated and has a smaller velocity dispersion. The more peaked features in the inner bins in \citet{Breddels2013}, and thus their radial anisotropy in the inner region, are likely to be caused by the combination of the two populations.  
The blue population dominates at $R>0.3$ kpc, so the tangential anisotropy we obtained for blue population is consistent with the result of \citet{Breddels2013} at this region. 

The single-component Jeans models by \citet{Walker2009jeans} and \citet{Lokas2009} are also consistent with highly-tangential velocity anisotropy; they could also be dominated by the features of the blue population, which is the dominant population and more spatially extended.

In conclusion, we find that the red population is more radially anisotropic (less tangential) than the blue population, which is consistent with the previous two-component models, while the tangential anisotropy of the blue population we obtained is consistent with the single-component orbit-based Schwarzschild model in the outer parts, where the blue population dominates.

In our model, the accuracy of the recovery of the velocity anisotropy is limited by the constant $\beta_z$ we assumed for each population. As shown in Figure~\ref{fig:kin_freekap}, the blue stars tend to have a higher degree of velocity anisotropy in the outer regions than the inner regions, which is not matched by our model perfectly. If we have more data points in the future, these features can be matched better by allowing $\beta_z$ to vary with radius. 

Tangentially-biased velocity anisotropy at large radii is a natural result of the dynamical evolution of a stellar system within an external tidal field, which induces a preferential loss of stars on radial orbits (e.g., \citealt{Takahashi&Lee2000}; \citealt{Baumgardt&Makino2003}; \citealt{Hurley&Shara2012}). The blue population of Sculptor is older and extends to much larger radii (see Section~\ref{S:results}), and, hence, is more likely to be tangentially biased by tidal forces.

\subsection{Internal rotation}
\label{SS:rotation}
\begin{figure}
\centering\includegraphics[width=\hsize]{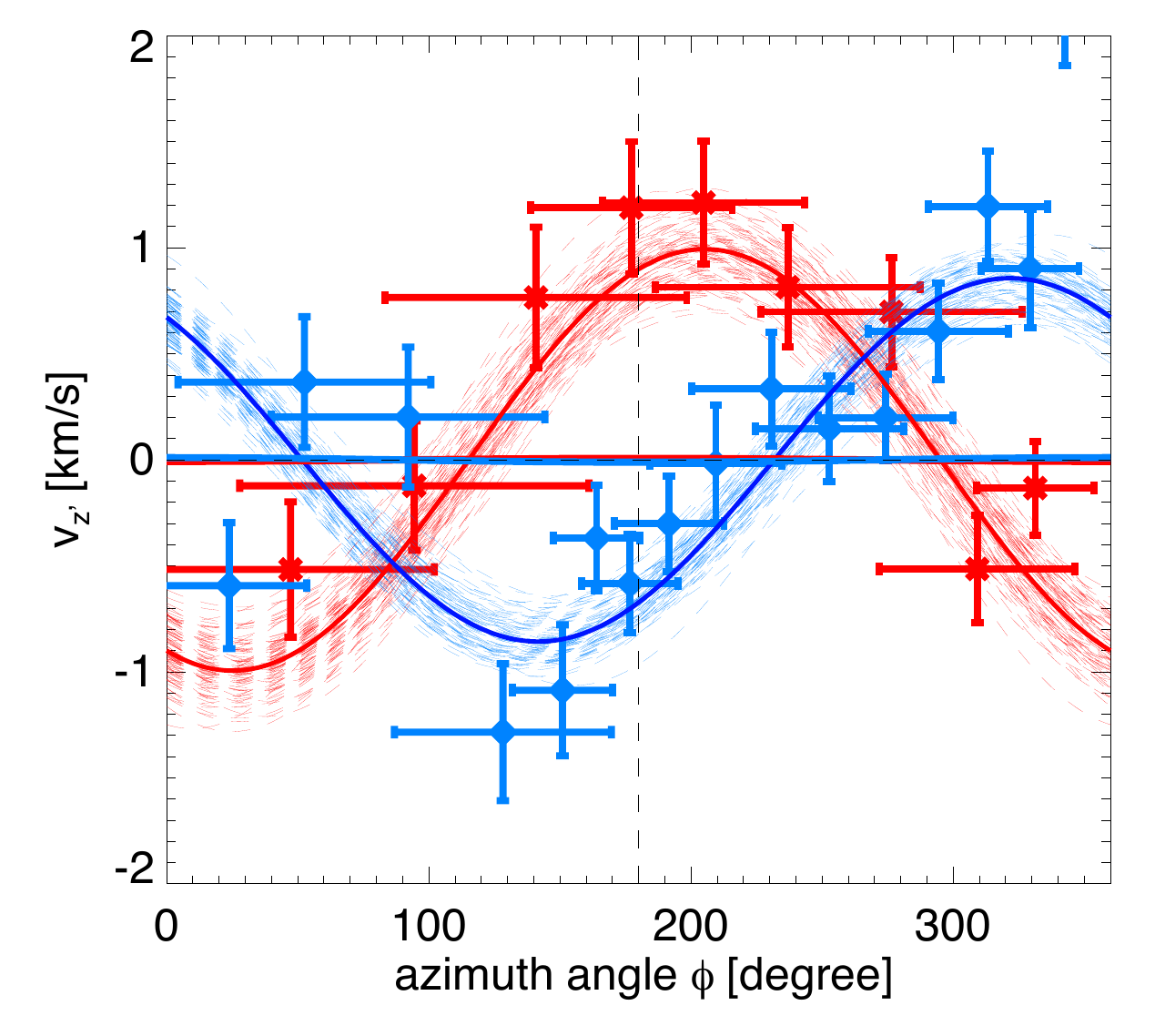}
\caption{The variation of line-of-sight velocity $v_{z'}$ along the azimuthal angle ($\phi$) from the
major axis of the galaxy. The red asterisks / blue diamonds represent the metal-rich/metal-poor population identified by the cored model with $\kappa$ free. The thin red/blue dashed lines are the direct sinusoidal fit to the data. The vertical dashed line indicates the position of $\phi = 180^o$. }
\label{fig:phivz_a1gf}
\end{figure}

The possible internal rotation of Sculptor was first discussed in \citet{Battaglia2008a}. 
We detect a possible rotation of the red population with $\kappa^{red} \sim 0.2$. 
The variation of line-of-sight velocity $v_{z'}$ along the azimuthal angle $\phi$, increasing from the major axis of the galaxy is shown in Figure~\ref{fig:phivz_a1gf}.  
Binning along $\phi$ is performed with 80 stars per bin and the bins close to each other have 40 stars of overlap. The red asterisks are the binned red stars and the blue diamonds are the blue stars, which are separated with the criterion of $P_i^{'red} > 0.6$ as red stars and $P_i^{'blue} > 0.6$ as blue stars. The cross contamination is more robustly removed in this separation, as the scatter in the rotation, especially for the red population, is smaller than that with stars separated with the criterion of 0.5 as used in previous sections. 
Simple sinusoidal fits of $v_{z'} = v_{max} \sin(\phi + \phi_0)$ to the binned data of the red stars yield the thin red dashed curves with $\phi_0 = 253^o\pm12^o$ and $v_{max} = 1.1\pm0.1$ \kms, which corresponds to $v_{max}/\sigma_0 = (1.1\pm0.1)/7.4 = 0.15\pm0.02$.  The maximum rotation occurs at $\phi \sim 0^o$ and $\phi \sim 180^o$, thus the rotation is around the minor axis.  For the blue stars, we get $\phi_0 = 160^o \pm 8^o$, so that the rotation is around a different axis. With $v_{max} = 0.9 \pm0.1$ \kms as indicated by the thin blue dashed curves, we get $v_{max}/\sigma_0 = (0.9\pm0.1)/10.6 = 0.09\pm0.01$.  
These stars are separated by their likelihood in the best-fitting models, the errors of $v_{max}/\sigma_0$ from the curve-fitting is small. 

However, stars are separated differently in different models, so the overall error of $v_{max}/\sigma_0$ is a combination of the statistical error from the MCMC process and the curve-fitting error, with the former one dominating. We randomly choose 100 models in the last steps of the MCMC process, identify the two populations by their likelihood in each of the model, and do the curve-fitting for the resulting red and blue populations.
We obtain $v_{max}/\sigma_0 = 0.15\pm0.15$ for the red population and $v_{max}/\sigma_0 = 0.09\pm0.15$ for the blue population. 

The binning reduces the fluctuation and increases the significance of the rotation. The reduced $\chi^2$ of the best sinusoidal fit to the red population is 1.8, while the reduced $\chi^2$ of the best fit to the blue population is 4.3. If we assume the blue stars have the same rotation as the red stars, the reduced $\chi^2$ of the blue stars will significantly increase to 27, thus indicating that the blue population is not well described by the rotation profile of the red population.  

The perspective rotation caused by global PM could be the same order as the rotation of the red population as we obtain here \citep{Walker2008} . However, the perspective rotation caused by the global PM should be the same for the red and blue population, which is not the case here. The rotation of the red population is thus likely to be true intrinsic rotation. 

Could the rotations be artificially introduced by the model? We further test this by creating a new model without rotation ($\kappa^{red} = \kappa^{blue} = 0$ fixed) as shown in Appendix~\ref{S:A2}, the rotation pattern of the red population still exists in the stars identified by the zero-rotation model, thus it is not likely to be artificially introduced. 

As we tested with the mock data, when there is such a weak intrinsic rotation in the red population, we could only recover it with $1\sigma$ significance, and the model is likely to slightly underestimate the rotation. Therefore, the real rotation in Sculptor may be stronger than we have estimated here.
An increased sample with $\sim 6000$ data points could significantly improve the statistical results. As the amplitude of the rotation is smaller than the typical velocity error of a single star, more accurate velocity measurements will also improve the inference of intrinsic rotation.

\section{Conclusions}
\label{S:conclusions}
We have presented a new chemo-dynamical modelling technique that separates multiple populations while simultaneously modelling their dynamics.  
This is achieved by extending the single-component discrete Jeans modelling of \citet{Watkins2013} to include multiple populations, each with different spatial, chemical and dynamical properties.
The probability of each star is a combination of its probability to be in either of the populations or to be part of the explicitly-modelled contamination.

We apply this modelling technique to the dSph galaxy Sculptor and find that:
\begin{itemize}
\item[-] We are able to identify the kinematics of multiple populations. The stars are naturally separated into two populations. The metal-rich (red) population is more spatially concentrated and has smaller velocity dispersion. 
\item[-] Assuming axisymmetry, a generalised NFW halo with inner density slope $\gamma$ left free converges to $\gamma = 0.5\pm0.3$, in between a core of $\gamma = 0$ and a cusp $\gamma =1$. We cannot exclude either the cored profile or the cusped profile at better than $1\sigma$ significance with the current data.
\item[-] The metal-rich population is nearly isotropic. The metal-poor population is close to isotropic in the inner regions and is moderately tangentially anisotropic in the outer regions. 
\item[-] We detect a $1\sigma$ significant intrinsic rotation of the red population with $v_{max}/\sigma_0 = 0.15 \pm 0.15$.
\end{itemize}

Our test using mock data show that to further constrain the inner density slope and the possible intrinsic rotation, we need more and/or better data points. A sample with $\sim 6000$ data points we are using could reduce the uncertainties by half and so address the `core' versus `cusp' problem under the assumption of axisymmetry.  

The discrete chemo-dynamical models that we have presented here are both powerful and flexible and can be applied to many other multiple-population systems for which discrete data are available. For example, giant elliptical (gE) galaxies usually have at least two populations of globular clusters (GCs) that are expected to have different dynamical properties due to their different formation histories. Not only will our method be able to separate the different GC populations, but its flexibility will allow us to include even more populations (such as planetary nebulae) and also to include integrated stellar kinematics from the inner regions. We have also applied our models to the gE galaxy NGC 5846 (Zhu et al., 2016), which shows that our method is able to simultaneously chemo-dynamically separate multiple populations, investigate the dynamical properties of multiple tracers, and constrain the underlying gravitational potential.

\section*{Acknowledgements}
The authors thank Nicolas Martin and Paolo Bianchini for useful discussions, and Eric Emsellem for providing his MGE Python code. Computer runs were mainly
performed on the MPIA computer clusters \emph{queenbee} and \emph{theo}.
This work was supported by Sonderforschungsbereich SFB 881 ``The Milky Way System" (subprojects A7 \& A8) of the German Research Foundation (DFG). 


\bibliographystyle{mn2e}
\bibliography{sculptor}

\clearpage

\appendix

\section{The definition of rotation in the Jeans models}
\label{S:Arotation}

\cite{Cappellari2008} adopted a rotation parameterisation,
\begin{equation}
\label{eqn:A1}
[\overline{v_{\phi}}]_j = \kappa_j ([\overline{v_{\phi}^2}]_j - [\overline{v_{R}^2}]_j )^{1/2}.
\end{equation}
for each Gaussian component $j$. The rotation of the model is then obtained by summing over the contributions from all Gaussian components:
\begin{equation}
\label{eqn:A2}
\nu \overline{v_{\phi}} = \mathrm{sgn}(\omega) \times  | \omega |^{1/2},
\end{equation}
with
\begin{equation}
\omega =   \nu \sum_{j =1}^{N} \mathrm{sgn}(\kappa_j) \times \kappa_j^2 \times ([\overline{v_{\phi}^2}]_j - [\overline{v_{R}^2}]_j )    
\end{equation}
where  $\mathrm{sgn}(x)$ indicates the sign of $x$.
The above definition assumes that $([\overline{v_{\phi}^2}]_j - [\overline{v_{R}^2}]_j ) > 0$ for all Gaussian components, thus the sign of $\kappa_j$ determines the rotation direction of that Gaussian component and directly contributes to $\omega$.

However, the variation of the velocity anisotropy $\beta_{z}$ combined with the properties of underlying potential may cause $([\overline{v_{\phi}^2}]_j - [\overline{v_{R}^2}]_j )$ to be either positive or negative. 
In this case, the followed calculations of rotation in equation (38) of \cite{Cappellari2008} (and also equation A59 and A60 in \citet{Watkins2013}) actually takes: 
\begin{equation}
\label{eqn:A3}
\omega =   \nu \sum_{j =1}^{N} \mathrm{sgn}(\kappa_j) \mathrm{sgn}([\overline{v_{\phi}^2}]_j - [\overline{v_{R}^2}]_j ) \times \kappa_j^2 \times |[\overline{v_{\phi}^2}]_j - [\overline{v_{R}^2}]_j |. 
\end{equation}
Thus the sign of $[\overline{v_{\phi}^2}]_j - [\overline{v_{R}^2}]_j $ also contributes to the rotation direction, and so $\kappa_j$ loses its control over the rotation direction in an inexplicit way. At the same time, the velocity anisotropy $\beta_z$ which affects the sign of $([\overline{v_{\phi}^2}]_j - [\overline{v_{R}^2}]_j )$, is involved in the determination of rotation direction, and thus becomes degenerate with $\kappa_j$.

With the old definition of rotation following \cite{Cappellari2008} and \citet{Watkins2013}, the best model, we get among a few sets of models with different $\kappa_j^\mathrm{red}$ and $\kappa_j^\mathrm{blue}$, is with $\kappa_j^\mathrm{red} = \kappa_j^\mathrm{blue} = 0.3$. 
Both cored and cusped halo models predict counter-rotation for the blue and red populations and the rotation directions of all Gaussian components of the red population are flipped compared to the sign of $\kappa_j^\mathrm{red}$, mostly due to its radial anisotropy ($\beta_z^{red} > 0$). 
Sometimes only the rotation directions of some Gaussian components are flipped compared to $\kappa_j$, which causes a smaller rotation in total or even a counter-rotation core in the model even with the same $\kappa_j$ given.

The models fit the data well in this case, although with a likelihood worse than we obtained in the main part of the paper. But, with the complicated coupling between velocity anisotropy parameters and rotation parameters, it may be that a poor set of rotation parameters $\kappa_j^\mathrm{red} $ and $\kappa_j^\mathrm{blue}$ were chosen. When we let $\kappa_j^\mathrm{red}$ and $\kappa_j^\mathrm{blue}$ be constant for all Gaussian components and free, the models can hardly converge to match the rotations because of the degeneracy between $\beta_z$ and $\kappa_j$.

Because of these disadvantages, in the paper, we chose to redefine the rotation with:
\begin{equation}
\omega =   \nu \sum_{j =1}^{N} \mathrm{sgn}(\kappa_j) \times \kappa_j^2 \times |[\overline{v_{\phi}^2}]_j - [\overline{v_{R}^2}]_j |. 
\end{equation}

Correspondingly, the mean velocities about the projected coordinates change, equation A59 and A60 in \citet{Watkins2013} become:
\begin{equation}
\label{eqn:Ivphi}
    I \, \overline{v_{\tau}} \left( x',y' \right) = 2 \sqrt{\pi G} \int_{-\infty}^{\infty}
        \mathcal{F}_{\tau} \times \mathrm{sgn}(\mathcal{G}) \times | \mathcal{G} |^{\frac{1}{2}}   \; \; \mathrm{d} z'
\end{equation}
with
\begin{equation}
\mathcal{G} =  \nu \sum_{j=1}^{N} \mathrm{sgn}(\kappa_j)  \kappa_j^2   \mathcal{G}_j 
\end{equation}
where $\tau$ represents $x^{\prime}$, $y^{\prime}$ and  $z^{\prime}$ and $\mathcal{F}_{\tau} = R f_{\tau}$ which remains the same as in \citet{Watkins2013}. The only thing that changes is:

\begin{equation}
\label{eqn:Gk}
\mathcal{G}_j = \Bigg\arrowvert  \int_0^1  \sum_{k=1}^{M} \frac{ \nu_j
        q_j \rho_{0k} \mathcal{H}_{k}(u) u^2 \mathcal{D}}{1 - \mathcal{C} u^2} \; \; \mathrm{d} u \Bigg\arrowvert,
\end{equation}
where N is the total number of luminous Gaussians of the tracer number density, M is the total number of potential Gaussians.
We take the absolute value of the integration in $\mathcal{G}_j $, and thus $\kappa_j$ regains its control over the rotation direction of each Gaussian component. The calculations become more expensive as $N$ integrations over $u$ will be needed inside the integration of $z'$ in equation~(\ref{eqn:Ivphi}).

\section{Models without rotation}
\label{S:A2}

\begin{figure}
\centering\includegraphics[width=\hsize]{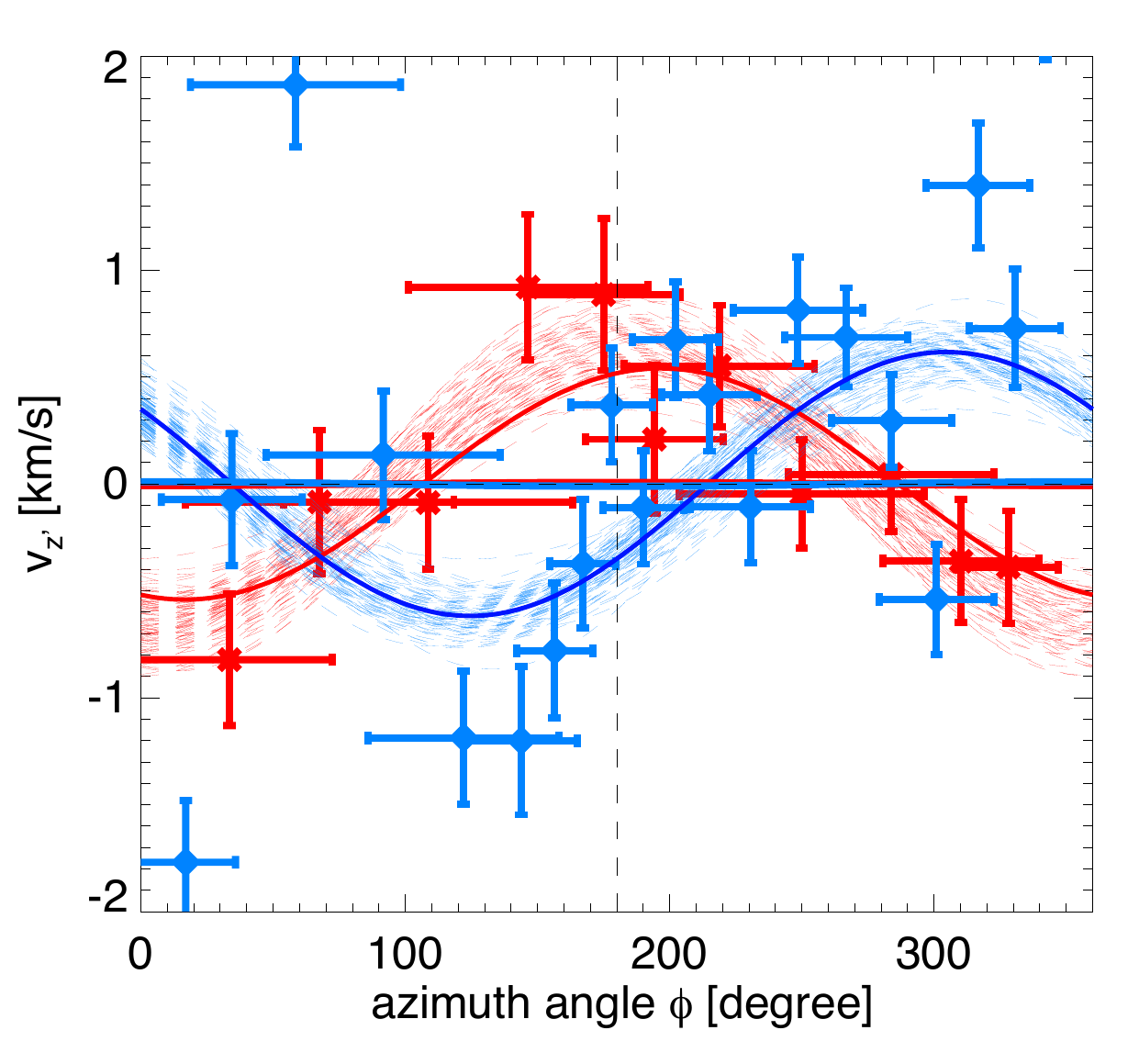}
\caption{The same as Figure~\ref{fig:phivz_a1gf}, but with the red and blue stars identified by the zero-rotation model. The same rotation pattern of the red population still exists, although with the amplitude suppressed by the zero-rotation model.}
\label{fig:a1g0_zerokap}
\end{figure}

In Section~\ref{SS:rotation}, we showed that the red and blue populations are counter-rotating when they are separated by the chemo-dynamical models with rotation.
To verify this finding, we also ran models without rotation (with $\kappa^\mathrm{red} = \kappa^\mathrm{blue}  = 0$ fixed). 
We separate the red and blue stars identified in this zero-rotation model as shown in Figure~\ref{fig:a1g0_zerokap}. 
Simple sinusoidal ($v_{z'} = v_{max} \sin(\phi + \phi_0)$) fits to the red stars yield the red curves with $\phi_0 = 268^o\pm12$ and $v_{max} = 0.8\pm0.1$ \kms, which corresponds to $v_{max}/\sigma_0 = 0.8/7.4 = 0.11$. The maximum rotation occurs at $\phi = 0^o$ and $\phi = 180^o$, thus the rotation is about the minor axis. 
The same rotation pattern as shown in Section~\ref{SS:rotation} exists in the red stars identified by the zero-rotation model. However, the zero-rotation model suppresses the possible rotations in each population, thus the amplitude of the rotation is decreased in the red population as identified by zero-rotation model. 

This test supports our results that the rotation pattern (at least in the red stars shown in Section~\ref{SS:rotation}) is not likely to be artificially introduced by our model. 

\bsp 

\label{lastpage}

\end{document}